\documentclass[aps,prb,reprint,superscriptaddress]{revtex4-2}

\usepackage{amsmath, newtxtext, newtxmath}
\usepackage{graphicx}

\usepackage{color}
\definecolor{bblue}{rgb}{0, 0.0, 0.8}
\definecolor{rred}{rgb}{0.7, 0.0, 0.0}
\definecolor{ggreen}{rgb}{0.0, 0.6, 0.0}
\usepackage{hyperref}
\hypersetup{
    colorlinks=true,        
    linkcolor=bblue,          
    citecolor=rred,        
    filecolor=bblue,      
    urlcolor=rred           
}

\begin{document}

\title{A concise 40 T pulse magnet for condensed matter experiments}

\author{Akihiko~Ikeda}
\email[E-mail: ]{a-ikeda@uec.ac.jp}
\author{Kosuke~Noda}
\author{Kotomi~Shimbori}
\author{Kenta~Seki}
\author{Dilip~Bhoi}
\author{Azumi~Ishita}
\author{Jin~Nakamura}
\author{Kazuyuki~Matsubayashi}
\affiliation{Department of Engineering Science, University of Electro-Communications, Chofu, Tokyo 182-8585, Japan}
\author{Kazuto Akiba}
\affiliation{Graduate School of Natural Science and Technology, Okayama University, Okayama 700-8530, Japan}
\affiliation{Faculty of Science and Engineering, Iwate University, Morioka, Iwate
020-8551, Japan}

\date{\today}

\begin{abstract}
There is a growing interest in using pulsed high magnetic field as a controlling parameter of physical phenomena in various scientific disciplines, such as condensed matter physics, particle physics, plasma physics, chemistry and biological studies.
We devised a concise and portable pulsed magnetic field generator that produces a 40 T field with a pulse duration of 2 ms.
It is assembled using only off-the-shelf components and a homemade coil that leverages small computers, Raspberry Pi, and Python codes.
It allows for straightforward modification for general purposes.
As working examples, we show representative applications in condensed matter experiments of magnetoresistance, magnetization, and magnetostriction measurements for graphite, NdNi$_{2}$P$_{2}$, and NdCo$_{2}$P$_{2}$, respectively, with the maximum magnetic field of 41 T and the lowest temperature of 4.2 K.
\end{abstract}

\maketitle

\section{Introduction}
A magnetic field is a critical control parameter in condensed matters that directly affects electron spin, electron's orbital angular momentum, the phase of electron's wavefunction, and nuclear spins in materials \cite{KartsovnikCR2004, TokuraRPP2006, ZapfRMP2014, LewinRPP2023}.
Not only that, researchers in diverse disciplines are investigating the magnetic field effects on vacuum fields \cite{BattestiPhysRep2018}, electromagnetic fluid dynamics in astrophysics \cite{ChibuezeNature2021, SakaiSR2022, MoritaPRE2022}, high-power laser plasma \cite{MoritaRMPP2023}, electrocatalytic reaction \cite{LiChemCatalysis2022}, the magnetoreceptivity of songbirds \cite{XuNature2021, BassettoNature2023, ReppertNature2024}, and magnetotactic bacteria \cite{BlakemoreScience1975, BlakemoreNature1980, UebeNatRevMicrobiol2016}.
The higher the magnetic fields get, the more evident the effect becomes.
In these ways, the demand for pulsed magnetic fields is growing in various scientific fields.

Obtaining high magnetic fields is technically demanding \cite{HerlachRPP1999}.
Superconducting magnets in laboratories provide DC magnetic fields of 20 T.
DC magnetic fields up to 45 T and pulsed high magnetic fields beyond 50 T are usually obtained only in large facilities for high magnetic fields.
Besides, localizing the magnets for each experiment is challenging, which poses a barrier to implementing high magnetic field experiments.
On the other hand, a pulsed magnet up to 40 T is obtained casually.
Small-scale pulse magnets can generate a maximum field of 50 T with a 100 times smaller pulse power system than those in large facilities. 
Small pulse magnets have been proven useful for the research of magnetic materials using x-ray diffractions \cite{YHMatsudaJPSJ2013, GerberScience2015, WenNC2019, WenNC2023}, x-ray spectroscopy \cite{NarumiPhysRevB2015, YamamotoPhysRevB2020, YamamotoPhysRevB2021, YamamotoPRB2024}, neutron scatterings \cite{OhoyamJPCS2006, YoshiiPRL2009, KnafoNC2016}. 
It has also been applied for the investigation of magnetic field effects on vacuum \cite{InadaPRL2017} and plasma \cite{AlbertazziScience2014}.
This technique has been mainly used for beam science because quantum beams require a relatively small magnetic field volume.

\begin{figure}[!hb]
\begin{center}
\includegraphics[width = 0.79\columnwidth]{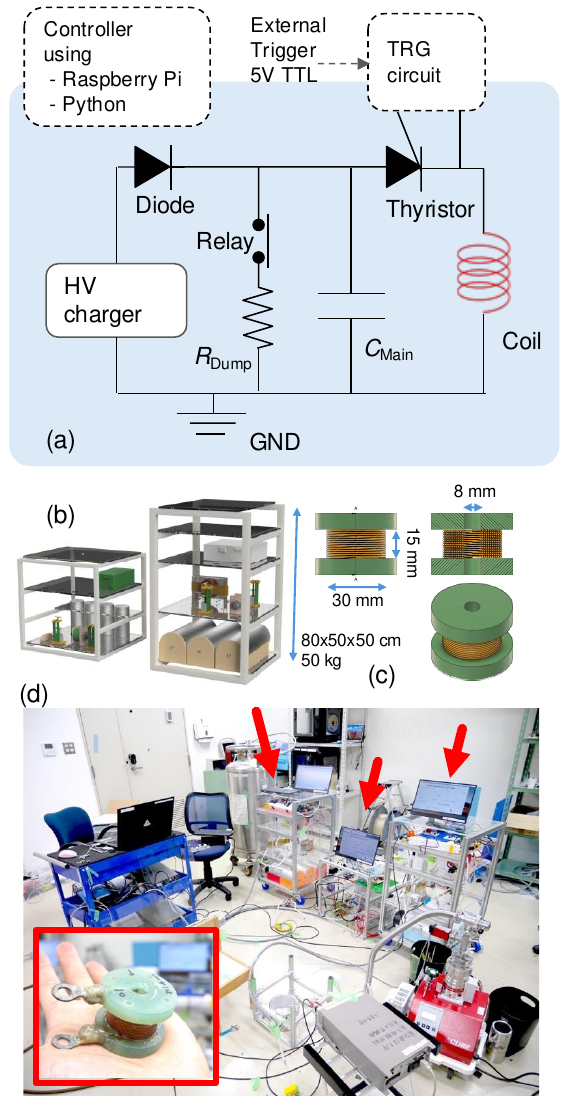}
\caption{
(a) Sketch of the pulse power circuit.
(b) Schematics of the assembled pulse power system. 
(c) The schematic drawing of a pulse magnet.
(d) A photo of our compact pulse-magnet-lab. with 3 assembled pulse powers. The inset is a photo of a homemade coil.
\label{fig1}}
\end{center}
\end{figure}

\begin{figure}
\begin{center}
\includegraphics[width = \columnwidth]{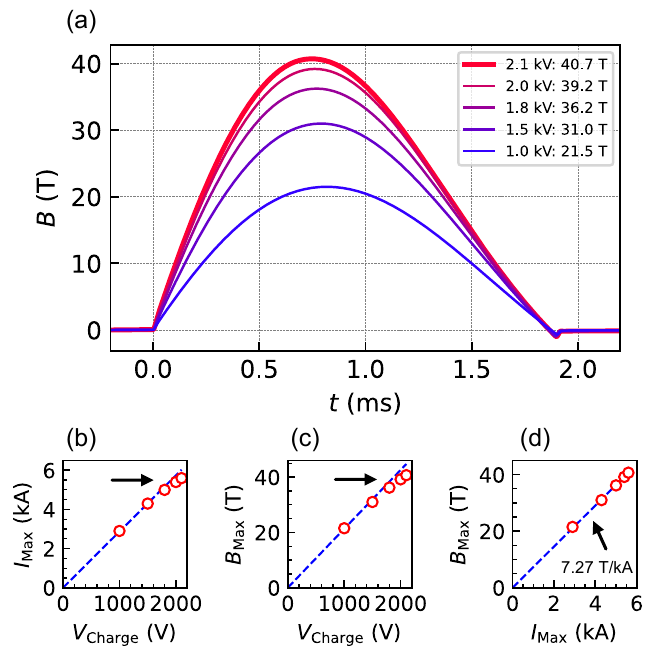}
\caption{
(a) Measured magnetic field profile as functions of time generated using the mini-pulse power and the mini-coil. 
(b) The relation of charging voltage and the maximum current. 
(c) The relation of charging voltage and the maximum magnetic field. 
(d) The relation of maximum current and maximum magnetic field. 
\label{fig2}}
\end{center}
\end{figure}

\begin{figure}[t!]
\begin{center}
\includegraphics[width = \columnwidth]{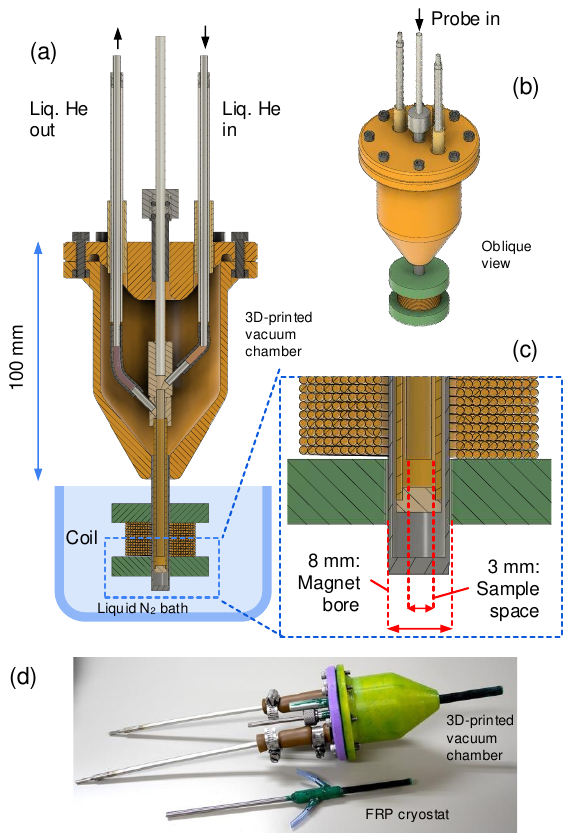}
\caption{
(a) A cross-section view of the He flow cryostat and the mini-coil. 
(b) Oblique view of the He flow cryostat and the mini-coil. 
(c) Magnification of the cross-section view of the cryostat and the mini-coil.
(d) A photograph of the 3D printed vacuum chamber and the FRP cryostat.
\label{fig3}}
\end{center}
\end{figure}

Small pulse magnets are even more promising for investing in broader scientific disciplines, including macroscopic laboratory experiments. However, such reports are missing.
Here, we show a portable pulse magnet system that weighs 50 kg in total, generating 40 T powered by RaspberryPi and Python codes.
It allows nearly everybody to have their own 40 T magnet in their laboratory, enabling rapid implementation of extremely high magnetic fields to new scientific fields.
As a working example, we show its application to condensed matter experiments with a good data quality.

\begin{figure*}
\begin{center}
\includegraphics[width = \textwidth]{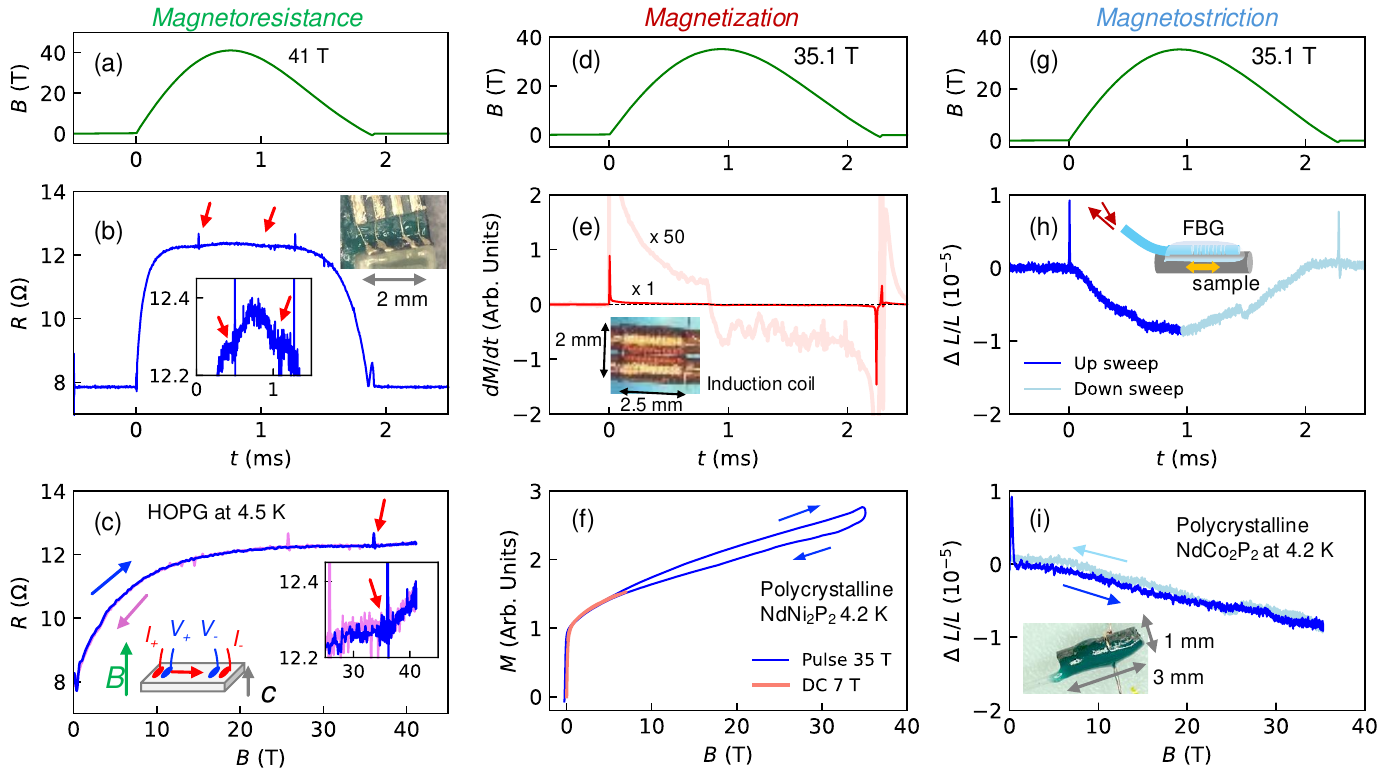}
\caption{
(a) Waveform of the pulsed magnetic field. 
(b) Magnetoresistance of HOPG as a function of time.
The inset on the left shows the magnification of the same plot where the CDW transition occurs. The inset on the right shows a photo of the HOPG sample with the 4 contacts for the 4 probe measurement.
(c) Magnetoresistance of HOPG as a function of the magnetic field.
The inset on the left shows a schematic drawing of the 4 probe measurement and orientations of the magnetic field, current, and crystal axis.
The inset on the right shows the magnification of the same plot where the CDW transition occurs.
(d) Waveform of the pulsed magnetic field.
(e) Measured signal of induction by $dM/dt$ of NdNi$_{2}$P$_{2}$.
The inset shows a photo of the inductive magnetization pick-up coil.
(f) Magnetization of NdNi$_{2}$P$_{2}$ as a function of the magnetic field.
(g) Waveform of the pulsed magnetic field, 
(h) Magnetostriction of NdCo$_{2}$P$_{2}$ as a function of time.
The inset shows a schematic drawing of magnetostriction measurement using FBG glued to the sample, 
(i) Magnetostriction of NdCo$_{2}$P$_{2}$ as a function of the magnetic field.
The inset shows a photo of the FBG and sample glued with SK-229 (Nitto).
\label{fig4}}
\end{center}
\end{figure*}

\section{Method}
The primary circuit involves three parallel capacitors (800 $\mu$F, 2000 V) to form $C=2.4$ mF (EPCOS: TDK Lambda), five times lighter than the conventional oil film capacitor.
The pulsed current is charged and discharged by a high-voltage charger (HUNS 3P117: Matsusada Precision), thyristor (5STP 17H5200: Hitachi ABB), Dump resistor, and high-voltage relay (Relay E12-NC: ROSS Engineering).
The thyristor's trigger circuit consists of a hand-made pulse transformer, optocoupler (FOD3182: Onsemi), and power MOSFET (2SK4017: Toshiba), which electrically separates external triggers from the high-voltage circuit.
The high voltage charger is controlled with variable DC voltage output (I$^{2}$C type DC10V output: NCD).
The high-voltage relay is controlled with a small relay array (Bit-trade one).
The charging voltage is monitored using an Analogue-Digital converter (I$^{2}$C type ADS1015: Analog devices).
All components are controlled by Python codes with Raspberry Pi, where the detailed Python codes are available in the Supplementary Note \cite{note1}.
A mini-coil is made from a round wire with a diameter $\phi$1 mm Cu-Ag alloy (Ag 10 \% in weight) and polyamide-imide enamel insulation.
We used a semi-auto coil winder made of a speed-control and torque-control motor.
The bore, winding number, and layers are 8 mm, 14, and 10, respectively.
The drawing of the circuit, image of the assembled system, and coils are shown in Fig. \ref{fig1}.

With the mono-coil, we obtained the highest magnetic field of 40.7 T at a current of 5.6 kA with a charging voltage of 2100 V as shown in Fig. \ref{fig2}(a), where the initial temperature of the coil is 78 K immersed in liquid N$_{2}$ bath.
As shown in Figs. \ref{fig2}(b) and \ref{fig2}(c), $I_{\rm{max}}$ and $B_{\rm{max}}$ as a function of charging voltage shows a deviation from the linear relation above 1800 V, indicating that the coil heats up by Joule heating resulting in higher resistance.
The result is in agreement with the result of the numerical calculation.

\section{Results}
As a working example of the pulsed high magnetic fields, we show the magnetoresistance, magnetization, and magnetostriction measurements for highly oriented pure graphite (HOPG), NdNi$_{2}$P$_{2}$, and NdCo$_{2}$P$_{2}$, respectively, in our laboratory.
To implement the low-temperature environment, we devised a plastic vacuum chamber and a plastic He flows cryostat \cite{TakeyamaJPE1988} as shown in Fig. \ref{fig3}.
Using this He flow cryostat, the sample temperature is variable from 300 K to 4.2 K by regulating the He gas flow.
First, we show the magnetoresistance measurement of HOPG at 4.5 K.
The electric conductivity is measured using the four-probe method with an AC excitation at 1 MHz and 1 mA.
The voltage at the sample is amplified using a pre-amplifier (SR445: Stanford research) and recorded with a digital oscilloscope (PicoScope 4284A).
The magnetoresistance of the sample is obtained by post-processing the recorded AC voltage of the sample using the numerical lock-in method \cite{MachelPhysicaB1995}.
The giant change in magnetoresistance with a saturating behavior is observed as shown in Figs. \ref{fig4}(b) and \ref{fig4}(c).
The kink features at 35.1 T, which is magnified in the insets of Figs. \ref{fig4}(b) and \ref{fig4}(c), indicate a charge density wave (CDW) transition, being in good agreement with previous reports \cite{FauquePRL2013, AkibaJPSJ2015}.

Next, we show a magnetization measurement of a ferromagnetic metal NdNi$_{2}$P$_{2}$ at 4.2 K up to 35.1 T, the moment of which is dominated by the local moment of Nd \cite{IshitaP}.
For measurement, the powder sample is fused in a Kapton tube installed in a parallel-type induction coil \cite{TakeyamaJPSJ2012}.
We observe the ferromagnetic moment as the initial ramp of magnetization followed by a gradual increase of magnetization with increasing magnetic field, as shown in Fig. \ref{fig4}(e) and \ref{fig4}(f).
The gradual increase is observed with high resolution, as shown in the magnified pink line in Fig. \ref{fig4}(e).
The observed open loop is not an intrinsic hysteresis of the sample but an error in the measurement.

Lastly, we show the result of magnetostriction of an antiferromagnet NdCo$_{2}$P$_{2}$ \cite{ReehuisJPCS1993, IshitaInteractions2024}  at 4.2 K in Fig. \ref{fig4}(h) and \ref{fig4}(i). 
The magnetostriction measurement uses fiber Bragg grating and optical filter method \cite{IkedaRSI2018, IkedaRSI2017}.
The results show a linear increase of magnetostriction up to 35.1 T, showing that a higher magnetic field is necessary for the magnetic saturation of the Nd moment. This indicates a relatively strong antiferromagnetic coupling between the Nd moment.

\section{Summary}
In summary, we show a pulse magnet with a simple design that weighs only 50 kg. 
It is easy to use and modify using Raspberry Pi and Python, whose detail is presented in the supplemental material \cite{note1}.
The application to condensed matter measurements is shown with good-quality data.
The concise pulse magnet system can provide broad scientific disciplines with pulsed high magnetic fields, such as particle physics, plasma physics, chemistry, and biology. We are also working on novel pulse magnets, such as a vector pulse magnet and AC 100 Hz pulsed high magnetic field.

\section*{Supplemental Material}
Supplementary Material shows how to build a 40 T generator, including the capacitor bank and a mini-coil.

\begin{acknowledgements}
Discussions with Y. H. Matsuda and H. Nojiri are acknowledged. 
This work is supported by MEXT LEADER program No. JPMXS0320210021, JST FOREST program No. JPMJFR222W, JSPS Grant- in-Aid for Scientific Research on Innovative Areas (A) 23H04861, 23H04859, 43 and Grant-in-Aid for Scientific Research (B) 23H01121.
\end{acknowledgements}

\section*{AUTHOR DECLARATIONS}
\paragraph*{\bf{Conflict of Interest}}
The authors have no conflicts to disclose.
\paragraph*{\bf{DATA AVAILABILITY}}
The data supporting this study's findings are available from the corresponding author upon reasonable request.

\bibliography{mini}
\end{document}


\title{Supplemental material}

\author{Akihiko~Ikeda}
\email[E-mail: ]{a-ikeda@uec.ac.jp}
\author{Kosuke~Noda}
\author{Kotomi~Shimbori}
\author{Kenta~Seki}
\author{Dilip~Bhoi}
\author{Azumi~Ishita}
\author{Jin~Nakamura}
\author{Kazuyuki~Matsubayashi}
\affiliation{Department of Engineering Science, University of Electro-Communications, Chofu, Tokyo 182-8585, Japan}
\author{Kazuto Akiba}
\affiliation{Graduate School of Natural Science and Technology, Okayama University, Okayama 700-8530, Japan}

\date{\today}

\begin{abstract}
In this Supplemental material, we will explain the overview and construction of a portable 40 T magnetic field generator and a tabletop 20 T magnetic field generator. 
\end{abstract}

\maketitle

\tableofcontents

\section{Design of Coils and Capacitor Banks for Specific Applications}
\subsection{Considering the Parameters of the Capacitor Bank Based on the Coil Shape and Target Magnetic Field}
In this section, we will design the coil and capacitor bank starting from the maximum magnetic field and the magnetic field volume. 
First, we consider the desired strength of the magnetic field. For now, we set a target magnetic field of 40 T, which can only be achieved with a pulse magnet without a special DC magnet in Tallahassee \cite{TothIEEE2018}.
Next, we consider how large the magnetic field volume is needed.
This also depends on the type of experiment in mind. Of course, a larger volume is better, but saving on magnetic field volume allows for a smaller magnetic field generation device. Here, we consider a cylindrical volume with a diameter and height of 1 cm. Then, we can calculate the energy stored in the magnetic field.
The energy in a magnetic field is given by
$E = \frac{B^{2}V}{2\mu_{0}}$
Calculating for 40 T, we get exactly 500 J. For 80 T, it becomes 2 kJ, for 400 T it becomes 50 kJ, and for 800 T it becomes 200 kJ. For those who prefer rounded values, 100 T corresponds to 3.1 kJ and 1000 T corresponds to 313 kJ. We are considering the case where all the energy stored in the capacitor, without losses, is converted into magnetic energy. In reality, this does not happen due to losses from Joule heating and stray fields. Estimating losses to be over 50 \%, let's set the capacitance to 800 $\mu$F.

\subsection{Determining LCR Circuit Parameters}
\subsubsection{Actual Calculation}
Now, we simulate the generation of a pulse magnetic field with a target of 40 T. The Python code is provided in \ref{simulation}. The calculation uses the Runge-Kutta method for numerical simulation of a standard LCR circuit, incorporating the temperature of the coil wire as a parameter. The inductance $L$ is calculated based on the coil dimensions using analytical formulas \cite{knoepfel1970}. The program is practical and roughly reproduces actual conditions under moderate scenarios, though it is not perfect, so use it with caution.

The initial calculation parameters are:
\begin{description}
\item[Capacitance:] 800 $\mu$F
\item[Magnet wire thickness:] $\phi$ 1 mm
\item[Bore diameter:] 1 cm
\item[Number of turns:] 10 turns
\item[Number of layers:] 10 layers
\item[Initial temperature:] 300 K
\end{description}
The residual $L$ of the capacitor bank is accounted for as 2.5 $\mu$H.

\begin{figure*}
\includegraphics[width= 0.8\textwidth]{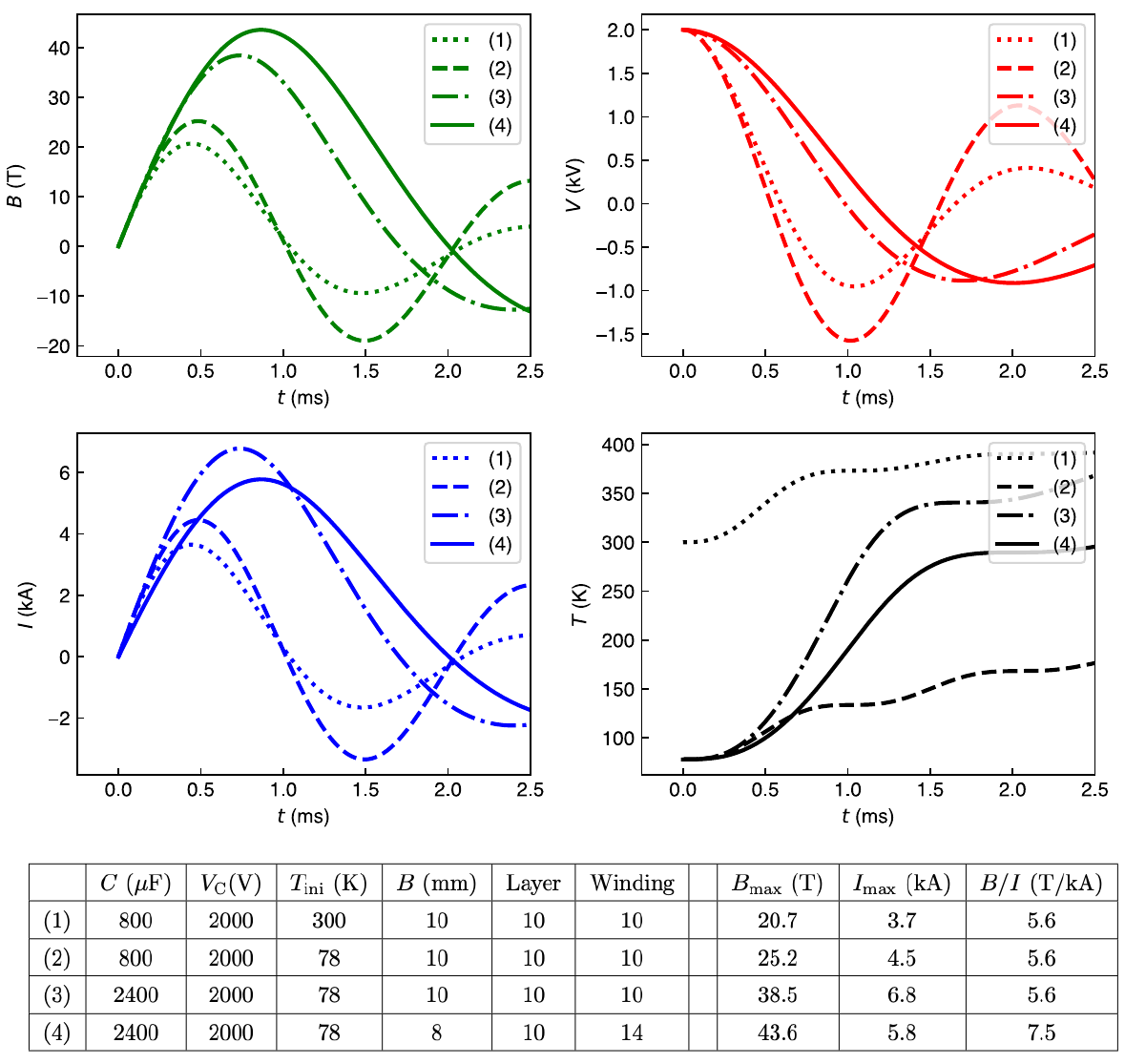}
\caption{Numerically calculated current ($I$) flowing through the load coil, generated magnetic field ($B$), voltage ($V$) across the load coil, and temperature ($T$) of the load coil, along with the parameters used for the calculation. \label{simresult01}}
\end{figure*}

The results from the first calculation are shown in Fig. \ref{simresult01}. The resulting magnetic field is 20.7 T with a pulse width of about 1 ms. The conversion efficiency from current to magnetic field is 5.6 T/kA. Various reasons could explain why it does not reach 40 T. First, note that the coil temperature rises to nearly 100 $^\circ$C. The coil uses not only copper wire and Kapton insulation but also epoxy and FRP (fiber-reinforced plastic). Since epoxy softens at high temperatures, rising to 100 $^\circ$C is problematic. Changing the initial temperature from 300 K to 78 K, let's perform the second calculation.

The magnetic field increases to 25.2 T. The increase in magnetic field is due to the rise in current. Because the initial temperature is lower, the increase in electrical resistance is suppressed, resulting in a current increase from 3.7 kA to 4.5 kA. The magnetic field efficiency, determined by the coil dimensions, remains unchanged. Although the temperature rises, it is within a tolerable range. Even so, it does not reach 40 T. Since it is possible to increase the capacitor, let's change the capacitance from 800 $\mu$F to 2400 $\mu$F and perform the third calculation.

The magnetic field value rises to 38.5 T, approaching 40 T. Since the coil dimensions are unchanged, this increase is due to the rise in current. The current rises to 6.8 kA. By tripling $C$, the pulse width increases from 1.0 ms to 1.7 ms, since $C$ affects the pulse width as the square root. The current should also increase as the square root of $C$, resulting in an actual increase from 4.5 kA to 6.8 kA, which is less than $\sqrt{3}$ likely due to increased electrical resistance from temperature rise. The temperature rises to 340 K, which is dangerous. Finally, considering increasing magnetic field efficiency, let's adjust the coil dimensions by changing the bore diameter from 10 mm to 8 mm and the number of turns from 10 to 14, and perform the fourth calculation.

The magnetic field value rises to 43.6 T, exceeding the target of 40 T. Additionally, the current decreases from 6.8 kA to 5.8 kA, resulting in a temperature decrease from 340 K to 290 K. These changes are due to modifications in the coil dimensions. Initially, the coil had $L = 130$ $\mu$H. Reducing the bore diameter to 8 mm increases the magnetic field efficiency from 5.6 T/kA to 6.1 T/kA, with an inductance of $L = 108$ $\mu$H. Extending the coil length to 14 mm increases the magnetic field efficiency to 7.5 T/kA, with an inductance of $L = 179$ $\mu$H. Consequently, a strong magnetic field is achieved at a lower current. The coil length could remain at 10 mm, but for reducing the current, consider designing for higher magnetic field efficiency. This results in increased $L$ and a broader pulse width. The current pulse width is close to 2 ms.

With numerous design parameters for coils and capacitor banks, it's essential to keep the final experiment in mind when considering coil design. Without setting appropriate limits on parameters, coil dimensions will never be finalized, preventing the start of capacitor bank design. For example, in this method, the coil length became longer. If planning a light scattering experiment, a longer coil would reduce the sample's viewing angle, decreasing signal strength. In such a case, extending the coil length would be undesirable. Moreover, reducing the bore diameter to 8 mm means a cryostat won't fit. Coil dimensions should be determined with the intended experiment and required magnetic field in mind. Some experiments need a smaller magnetic field volume, allowing for higher magnetic field efficiency by reducing the bore diameter to a few millimeters.

Human nature desires a broader, longer, and stronger magnetic field. Calculating various scenarios quickly shows that capacitor capacity, coil mass, and required voltage can increase tenfold or a hundredfold. Such strong magnetic fields are reserved for specialized facilities. Conversely, halving a parameter can shrink the scale of the capacitor bank and coil to 1/10, making it manageable even in individual labs or for retired professors at home. This greatly expands the potential, highlighting the value of mini banks and mini coils, emphasizing the importance of effective subtraction in mini bank fabrication.

For reference, the energy used in the fourth calculation parameter set (4) is 4.8 kJ. The magnetic energy when generating 40 T in a cylinder with a diameter of $\phi$ 1.0 cm and a length of 1 cm should have been 500 J. The bore volume decreased by 90 \%, resulting in an efficiency of 10 \%, raising doubts about its adequacy. This poor efficiency is partly because the coil deviates significantly from an ideal solenoid. An infinitely long coil with infinite layers would concentrate the magnetic field entirely within the bore, converting all the capacitor's charge energy into magnetic energy. However, in practice, increasing turns raises $L$, reducing the current. To forcibly increase the current, reducing $C$ and raising the voltage enlarges both the power supply and coil due to insulation requirements.

\subsection{Design of the Main Circuit}

\begin{figure*}
\includegraphics[width= 0.8\textwidth]{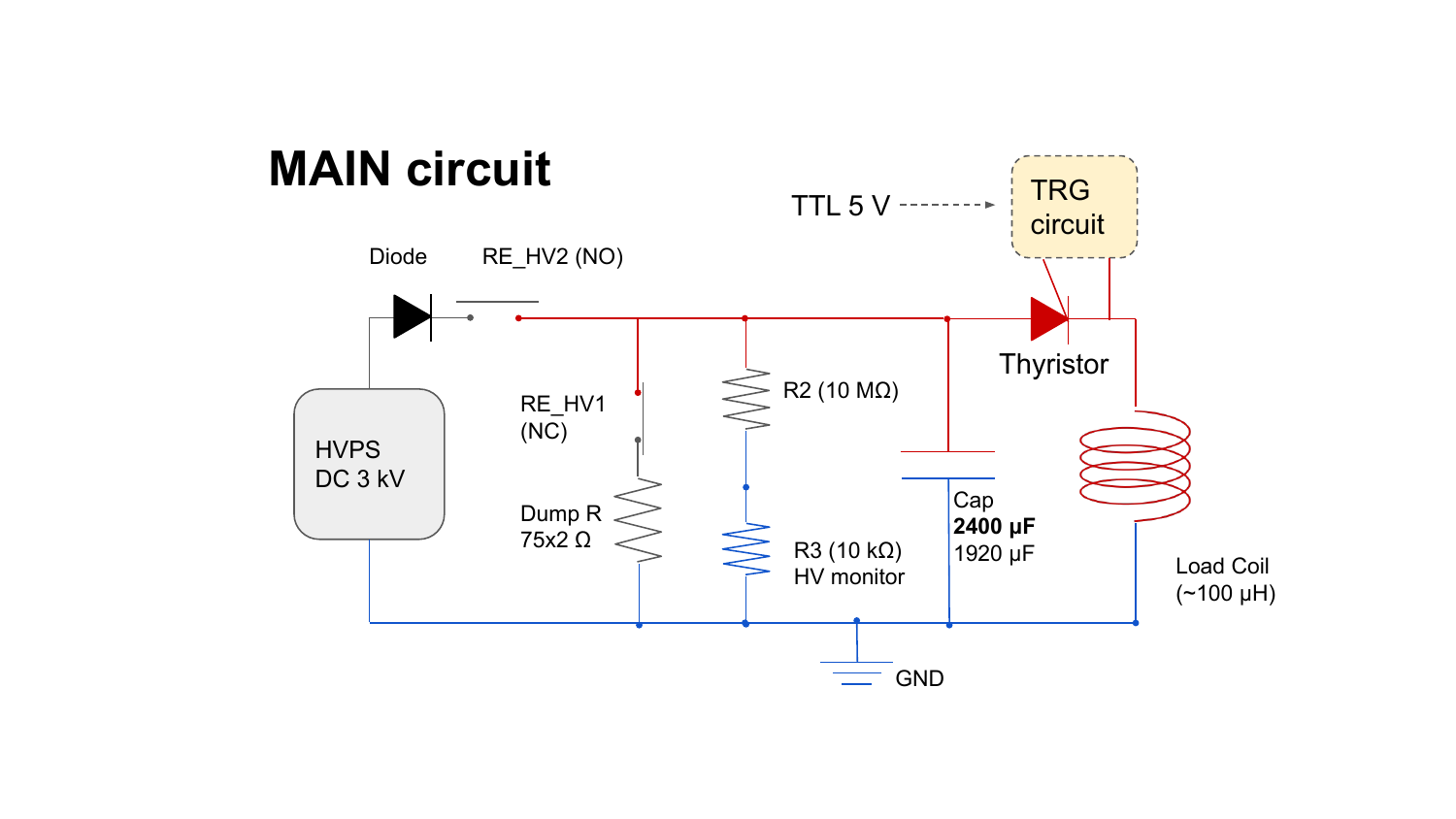}
\caption{Schematic diagram of the main circuit\label{maincircuit}}
\end{figure*}

We have discussed the relationship between $L$ and $C$ and the generated magnetic field. Let's proceed to create the main circuit. The main circuit diagram of the pulse circuit to be produced is shown in Fig. \ref{maincircuit}. The Cap is the main capacitor, and the output is connected to the load coil. There is a semiconductor switch called a Thyristor between them, which remains insulated until it is turned on. The Thyristor is a so-called power semiconductor, driven by an external trigger circuit. It is convenient to make the trigger circuit driven by a 5V TTL signal from an external source. These trigger circuits will be described later.

Immediately to the left of the capacitor are the voltage divider resistors R2 and R3, used to measure the voltage across the capacitor. R2 is a high-voltage high resistor, and by measuring the voltage of a few volts across R3, the voltage of several kV across the capacitor can be measured. Further to the left are the dump resistor and the grounding relay (RE HV1), which can be considered the lifeline of this high-voltage circuit. These are installed to safely and quickly discharge the electricity stored in the capacitor. The selection criteria for the dump relay include reaching 0 V within a short time of about 1 second and having sufficient heat capacity to absorb the Joule heat. This time, we used a module that connects two 75 \si{\ohm} resistors in series. The high-voltage relay needs to switch the high-voltage circuit on and off. It must be highly reliable and stable, with no room for malfunction. The high-voltage relay operates when it receives power from an external source. NC stands for Normally Closed, meaning that when there is no power, the switch is closed, and the circuit is conductive. When power is applied, the switch opens, making the circuit open. This ensures that in the event of a sudden power failure, the system will automatically ground itself. The switch used here must absolutely be NC. Until charging is to be performed, do not supply power to the dump relay. As a result, the red and blue parts of Fig. \ref{maincircuit} are all grounded.

Further to the left, a high-voltage DC power supply is installed as the charger. HVPS stands for High Voltage Power Supply. A switching power supply is used because it is lightweight and easily controllable. A charger that outputs several kV in response to an external 0-10 V analog voltage is chosen. Between the charger and the capacitor are a relay (RE HV2) to disconnect the charger and a protective diode. Both of these are to protect the charger. During pulse discharge, when the current reaches zero after a half-cycle, the capacitor becomes reverse-charged. If the blue part is considered GND and remains at zero volts, the red part becomes high negative voltage when reverse-charged. If the charger remains connected, it may cause a failure. To prevent this, the diode is grounded to prevent negative voltage, or the circuit is mechanically disconnected during pulse discharge. Either one should be sufficient, but there is no problem with having both. Either one can be omitted.

Finally, let's explain the basic operation of charging and discharging. The operation is as follows:
\begin{enumerate}
\item[]{Basic operation of charging and discharging}
\item Open RE-HV1 (dump relay)
\item Close RE-HV2 (HVPS relay) and start charging
\item When charging is complete, open RE-HV2 to stop charging
\item Supply an external TTL trigger to the trigger circuit to turn on the thyristor and discharge
\item Immediately after discharge, close RE-HV1 to return to the initial state
\end{enumerate}
If you look at this and think it makes sense, you roughly understand the basics of the circuit. If there are any unclear parts, please refer to Fig. \ref{maincircuit} and think it through step by step. The important point is the voltage at each part. In the initial state, since RE-HV1 is closed, almost all places, including the load coil, are grounded. This is a safe state. From step 1, when RE-HV1 is opened, the safety device is removed. It is correct to feel a sense of tension from here. In step 2, the entire high-voltage side of the capacitor becomes high voltage. This high-voltage state lasts for a long time, requiring insulation for DC high voltage, and careful attention is needed. Remember that in this state, the load coil remains grounded. When step 3 is completed, the pulse circuit becomes an isolated circuit. It is important to be able to quickly close RE-HV1 and return to the initial state if you decide to stop the experiment or if an abnormality occurs, such as a power failure. Let's consider the voltage distribution at the moment of step 4. In fact, the red areas in Fig. \ref{maincircuit} become high voltage, and the blue areas become GND. The red areas at this moment are called pulse high voltage or pulse high. The blue areas are called pulse ground. The locations grounded in a series of operations change moment by moment. Understanding which part has which voltage at which timing is crucial for the safe operation of the device.

\subsection{Thyristor Trigger Circuit}

\begin{figure*}
\begin{center}
\includegraphics[width= 0.8\textwidth]{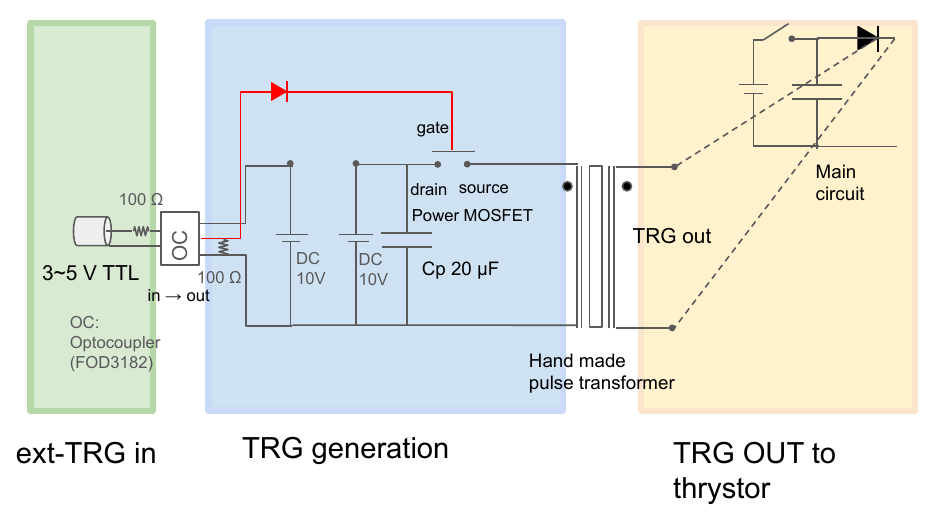}
\caption{Schematic diagram of the trigger circuit\label{trg}}
\end{center}
\end{figure*}

Following the main circuit, let us explain the thyristor trigger circuit. The thyristor starts when a specified current and voltage are applied. The required current and voltage depend on the thyristor, and it is necessary to read the specifications of the purchased thyristor. This will be described later. A thyristor starts with power; it does not start with voltage application alone. The basic characteristics of a thyristor are summarized as follows:
\begin{itemize}
\item The trigger requires both current and voltage (i.e., a certain amount of power).
\item It has rectifying action even after starting.
\item Once it is on, it remains on as long as current is flowing.
\item It turns off when the current reaches zero or when reverse voltage is applied.
\end{itemize}

The schematic diagram of the trigger circuit is shown in Fig. \ref{trg}. To trigger the thyristor, consider the load side as ground and apply positive voltage and current to the thyristor's trigger electrode. Since the trigger circuit is directly connected to the main circuit, the entire circuit becomes high voltage at the moment the thyristor is turned on. The problem is that the trigger circuit must also accept triggers from the outside. If everything is connected with DC, high voltage could reach the external triggering equipment, leading to failure. Therefore, a pulse transformer and optocoupler are used for double insulation (Fig. \ref{trg}). By using a pulse transformer, power is transmitted, but it is insulated from DC. It is necessary to use a pulse transformer with a sufficiently high insulation withstand voltage between the primary and secondary sides. This time, we used a homemade pulse transformer with high insulation voltage film wire. Additionally, an optocoupler is used to isolate the trigger input from the trigger generation circuit\footnote{We used an optocoupler (FOD3182) suitable for our conditions, recommended by Dr. T. Takezaki from Toyama University. Thank you.}. With double insulation, the trigger generator and measuring instruments can be protected. The differently colored areas in Fig. \ref{trg} are electrically insulated, allowing the safe transmission of trigger pulses.

Next, let's create a pulse circuit on the primary side of the pulse transformer. For reasons described later, we will charge a capacitor of about 20 $\mu$F with about 10 V and create a pulse discharge circuit. The basic principle is the same as the main circuit, but since the current and voltage are only a few amperes and a few volts respectively, it can be easily prototyped and tested. Use a small film capacitor for the capacitor and a power MOSFET for the switch. First, input a 3-5 V TTL signal externally to the optocoupler. Connect the output of the optocoupler to the gate of the MOSFET. Use a DC power supply to generate 10 V. This circuit is compact and can be made on a breadboard.

\section{Selection, Implementation, and Procurement of Components}
\begin{figure}
\includegraphics[width=\columnwidth]{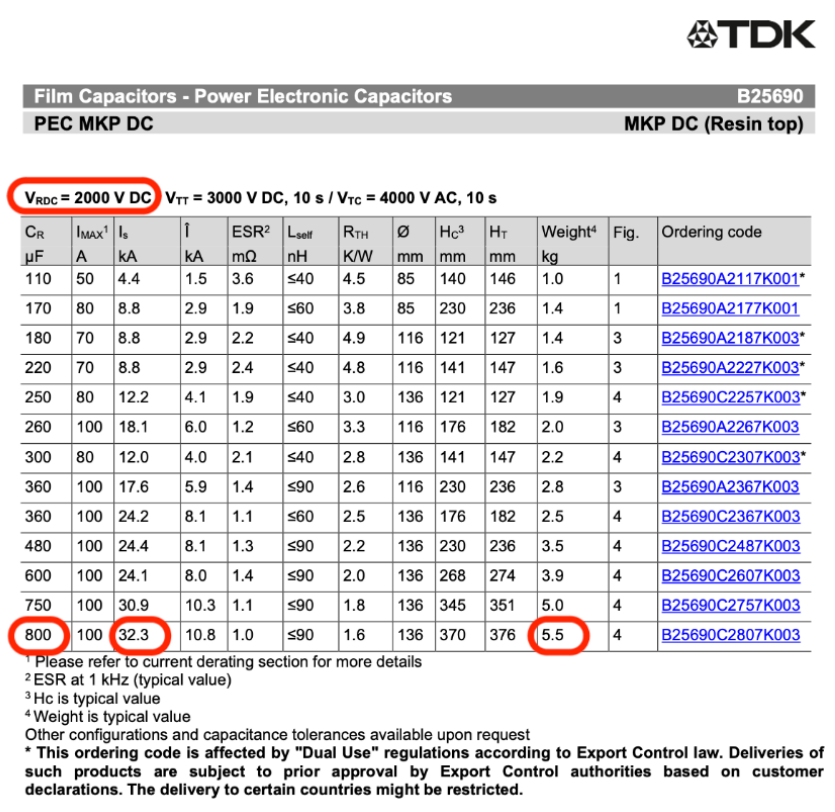}
\caption{Example of capacitor specification sheets and key parameters to note\label{spec_caps}}
\end{figure}
Up until now, we have been dealing with conceptual explanations. From here on, we will describe the selection method of the actual parts required for implementation and how to implement them.

\subsection{Selecting Components for the Main Circuit}
\subsubsection{Selecting a Capacitor}
The specifications to read when selecting a capacitor are the DC charging voltage ($V_{\rm DC}$) and the surge current value ($I_{\rm S}$). $V_{\rm DC}$ is the maximum voltage that can be constantly charged. $I_{\rm S}$ is the maximum instantaneous current that can flow. An example of a capacitor specification sheet is shown in Fig. \ref{spec_caps}.

Recently, dry film capacitors have become available\footnote{Professor H. Nojiri from Tohoku University IMR mentioned, "There's something like this, do you think it can be used?" So I tried using it.}. Let's take an EPCOS (TDK lambda) 800 $\mu$F at 2 kV capacitor as an example. Compared to traditional oil film capacitors, the dry film capacitors used this time are much more space-efficient and significantly lighter, which is advantageous for creating compact devices. The surge current value of traditional oil film capacitors with the same capacity is 6 kA, whereas the dry film capacitors have a surge current value of 32.3 kA, more than five times larger per unit capacity. They also weigh 5.5 kg, about 1/4 the weight of traditional oil film capacitors per unit capacity. Although the durability is uncertain, there have been no problems so far. Additionally, the cost is lower. The price is less than 1/4 of traditional oil film capacitors. However, it is heard that for higher voltage products, traditional oil-filled film capacitors are still advantageous. The dry film capacitors used this time are readily available from sites like Digikey and RS Components, making them easy to procure.

\subsubsection{Selecting a Thyristor}
\begin{figure}
\includegraphics[width=\columnwidth]{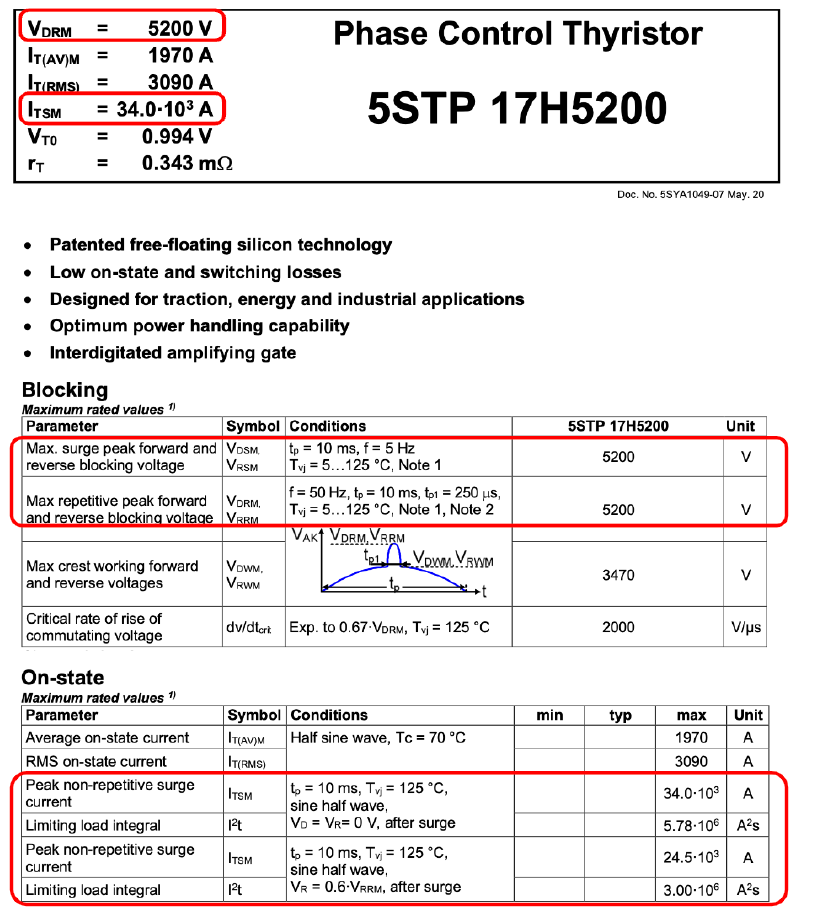}
\caption{Example of thyristor specification sheets and key parameters to note\label{spec_thyristor}}
\end{figure}

The heart of the circuit is the thyristor. The primary specifications to look at for a thyristor are the maximum forward voltage $V_{\rm M}$ and the maximum forward surge current $I_{\rm TSM}$. An example of a thyristor specification sheet is shown in Fig. \ref{spec_thyristor}.

The maximum forward voltage indicates how much voltage can be applied to switch from off to on. Additionally, the maximum forward surge voltage, maximum reverse voltage, and maximum reverse surge voltage may also be listed. These are all voltages that can be applied to the switch either continuously or instantaneously. Since a voltage is always applied to the thyristor during charging, the forward voltage needs to be sufficiently higher than the charging voltage. The reverse voltage is particularly important. When the thyristor turns on, voltage reflections may occur in the coil, momentarily applying a high reverse voltage. Therefore, it is also necessary to consider the reverse voltage. It is safer to choose a slightly over-specced thyristor to prevent damage.

The maximum forward surge current $I_{\rm TSM}$ determines the maximum current that can flow through the coil and is the most important value. Often, the pulse width and temperature conditions for the maximum current are also listed. For instance, it might be specified that 10 kA can flow, but this assumes a 10 ms pulse. This implies that for a 1 ms pulse, more current can flow. A rough calculation method involves keeping $I_{\rm TSM}^{2}\Delta t$, which is related to the amount of heat generated, constant. For example, consider a case where $\Delta t = 10$ ms is specified, and we assume $\Delta t = 1$ ms. Multiplying $I_{\rm TSM}^{2}$ by 10 implies that $I_{\rm TSM}$ can flow up to about 3.16 times more. If a thyristor fails, it cannot be repaired and must be replaced.

\begin{figure}
\begin{center}
\includegraphics[width=0.9\columnwidth]{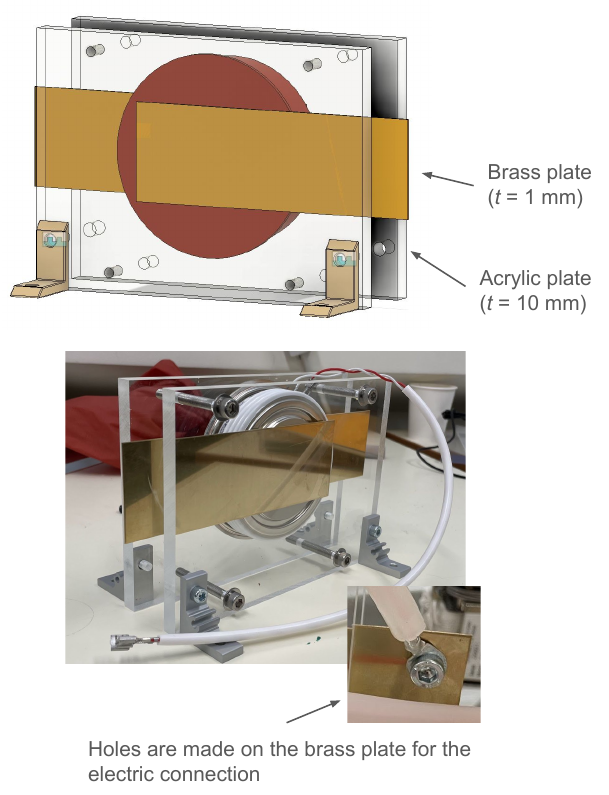}
\caption{Example of a prototyped thyristor module assembly\label{thyristor_module}}
\end{center}
\end{figure}

\begin{figure}
\begin{center}
\includegraphics[width=\columnwidth]{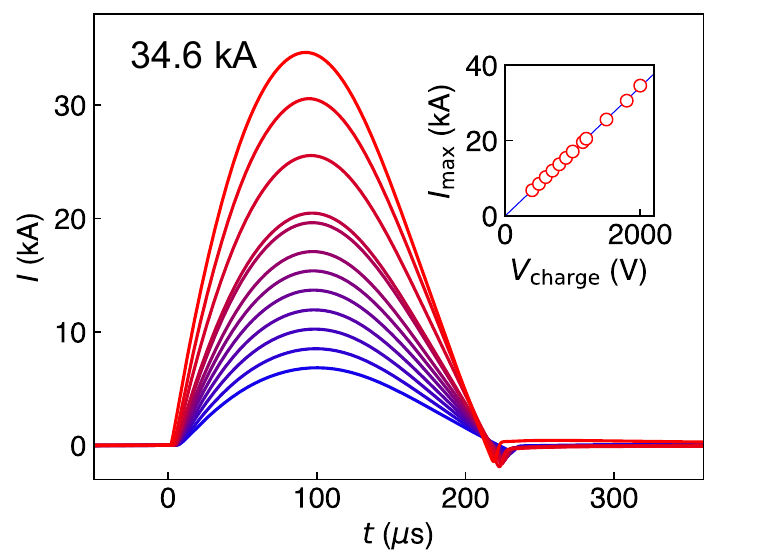}
\caption{The waveform of the test discharge using the thyristor up to 34.6 kA (load shorted).\label{testdischarge}}
\end{center}
\end{figure}

For the 40 T device, we selected a thyristor from Hitachi ABB. Hitachi ABB offers a wide range of high-performance products, demonstrating high technical capability. This time, we chose $V_{\rm M} = 5.5$ kV and $I_{\rm TSM} = 34$ kA\footnote{Note that the delivery time for this thyristor has been at least eight months in recent years.}. The bulk thyristor module requires custom electrodes for connections. An example of the thyristor module we created is shown in Fig. \ref{thyristor_module}. The result of the test discharge is shown in Fig. \ref{testdischarge} where a high current of 34.6 kA is successfully generated.

For smaller-scale devices with lower voltage and current, inexpensive thyristor modules can be used. These are $V_{\rm M} = 1.8$ kV and $I_{\rm TSM} = 2$ kA (10 ms) from Semikron. Assuming a 1 ms pulse, it can be assumed that up to about 6 kA can flow, which is sufficient for a 20 T device. Other companies like IXYS and VISHAY offer similar products, all of which are considered equivalent. These can be purchased from Digikey or RS Components. These modules come with mounting screw holes for the electrodes and holes for fixing the device itself, allowing them to be used as is.

\subsubsection{Selecting a Charger}

Use a lightweight switching power supply for the charger. This time, we selected a model that can control output with an external analog voltage up to 10 V and can start and stop charging using an open collector. Choose the output wattage so that the charging time is about one minute. Calculate the number of coulombs to be stored in the capacitor and divide it by the maximum current to find the shortest charging time. For example, for 2 kV and 2.4 mF, $Q = 4.8$ C. A power supply that can charge at 120 mA will charge in 40 seconds, but at 40 mA, it will take three times longer, about 2 minutes. However, a power supply with lower wattage is smaller, so if miniaturization is the priority, choose the one with lower wattage. With a voltage of about 2 kV, there is little danger even if it takes a long time. Many domestic and international manufacturers handle DC chargers, so any will do. This time, we chose the HUNS series from Matsusada Precision.

\begin{figure*}
\includegraphics[width=\textwidth]{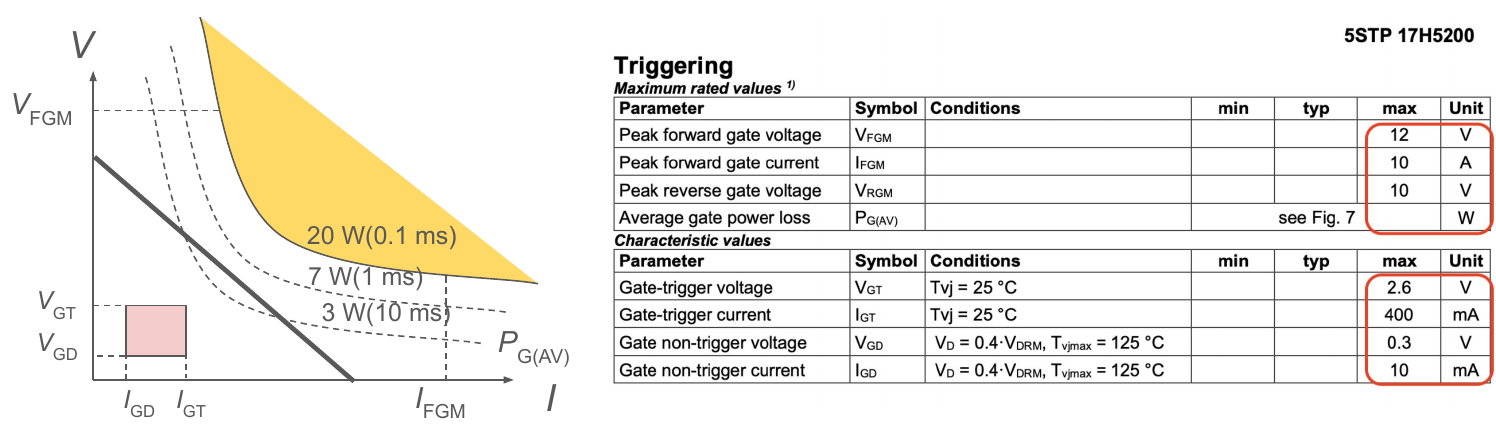}
\caption{Relationship between thyristor trigger current and voltage, example of a thyristor trigger parameter sheet, and key parameters\label{trigger}}
\end{figure*}

\subsubsection{Other Components}
For high-voltage relays used in the main circuit, reliable ones from Ross Engineering are recommended. The available ones are quite over-specified, starting at 12 kV. Each costs around 100,000 yen. One NC relay is absolutely necessary, but NO relays can be omitted by installing high-voltage protection diodes. I have a bitter memory of once forgetting to open the NO relay before discharging, which broke the charger.

For high-voltage protection diodes, they should be able to carry the charging current and have a sufficiently high blocking reverse voltage. For the charger prepared this time, the maximum forward current is 120 mA, and the blocking reverse voltage is 2 kV. If this protection diode fails to work, the charger will break, leading to a sad experience. Therefore, we prepared a 10 kV, 1 A diode.

For the grounding resistor, which absorbs the energy, prepare a large donut resistor. Connect two 75 \si{\ohm} resistors in series to make 150 \si{\ohm}. With a 2 kV capacitor, the initial current is calculated to exceed 10 A. Assuming 4.8 C from the previous calculation, it is expected to reach zero volts in about one second. Dividing the capacitor energy of 4.8 kJ by the heat capacity, the temperature rise can be estimated, but it is only a few kelvins, almost unnoticeable. Even touching it, you wouldn't feel it, making it sufficient.

\subsection{Design and Component Selection for the Thyristor Trigger Circuit}

To prepare a trigger for the thyristor, it is necessary to understand the thyristor's specification sheet. There are many items to read, which are listed below. These terms relate to the current and voltage applied between the trigger gate and the cathode.
\begin{enumerate}
\item $V_{\rm GT}$: Gate Trigger Voltage
\item $V_{\rm GD}$: Maximum Non-Gate Trigger Voltage
\item $I_{\rm GT}$: Gate Trigger Current
\item $I_{\rm GD}$: Maximum Non-Gate Trigger Current
\item $V_{\rm GTF}$: Maximum Peak Gate Forward Voltage
\item $I_{\rm GTF}$: Maximum Peak Gate Forward Current
\item $P_{\rm G(AV)}$: Allowable Gate Power Dissipation
\end{enumerate}
It is important to correspond these parameters to Fig. \ref{trigger} (left figure) and ultimately be able to draw an oblique line. Let's explain them in order.

Using the first four parameters, we can draw a rectangle in the lower left of the graph on the left side of Fig. \ref{trigger}. The switch will not operate in the region below and to the left of this rectangle. Within this rectangle, the switch may or may not operate, leading to unstable operation. The specification sheet in Fig. \ref{trigger} lists the maximum non-gate trigger voltage $V_{\rm GD}$ and the maximum non-gate trigger current $I_{\rm GD}$ as 300 mV and 10 mA, respectively. These values are not large. If a current and voltage larger than these values are input to the thyristor's trigger gate, the thyristor might start. This indicates that small noise from the surrounding environment could cause the thyristor to malfunction. This shows the characteristic of thyristors that they can start with a slight trigger voltage and current, making them prone to malfunction. Above and to the right of the rectangle is a region where the thyristor will always operate. However, there are limits to the current, voltage, and power. The limit on power is shown by the allowable gate power dissipation $P_{\rm G(AV)}$. This condition indicates that the gate will be damaged if power beyond the specified amount is input, with the limit varying based on the duration of the power. For example, if a trigger pulse of about 0.1 $\mu$s is applied, the allowable power is 20 W, represented as the yellow region in the left graph of Fig. \ref{trigger}. Furthermore, the maximum peak gate forward voltage $V_{\rm GTF}$ and the maximum peak gate forward current $I_{\rm GTF}$ indicate the limits for voltage and current, respectively.

\begin{figure}
\begin{center}
\includegraphics[width=\columnwidth]{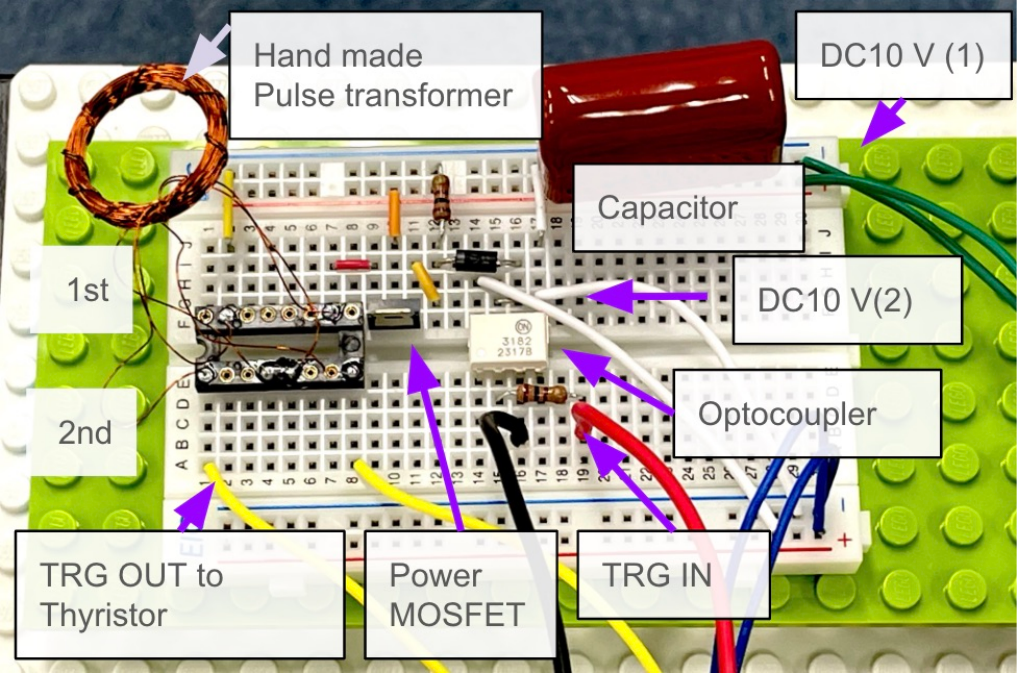}
\caption{Photo of a home-made trigger circuit\label{trigger_photo}}
\end{center}
\end{figure}

So, how do we safely apply current and voltage to the gate electrode? If the gate resistance is fixed, the current and voltage will be uniquely determined once one is set. However, the gate resistance dynamically changes from open to low resistance, making it difficult to predetermine. The maximum voltage and current can actually be determined by the following method. This way, regardless of the gate resistance, the current and voltage values will lie on the thick diagonal line in the left graph of Fig. \ref{trigger}. This allows the thyristor trigger circuit to operate stably. First, recall that the trigger circuit is also a pulse discharge circuit, as shown in Fig. \ref{trg}. Although a pulse transformer is installed to isolate it from the main circuit in Fig. \ref{trg}, let's consider the case where the pulse circuit is directly connected to the thyristor without the transformer. Assume the thyristor's gate resistance is infinite when it is off. Then, the voltage across the gate equals the capacitor's charging voltage. This determines the maximum voltage, intersecting the thick diagonal line and the $V$ axis in the left graph of Fig. \ref{trigger}. Next, consider when the thyristor's gate resistance becomes zero. By appropriately selecting the gate series resistor $R_{\rm GS}$, the maximum current can be determined along with the capacitor voltage. This determines the intersection of the thick diagonal line and the $I$ axis in the left graph of Fig. \ref{trigger}. The circuit design involves setting the capacitor's charging voltage and the gate series resistor $R_{\rm GS}$ so that this diagonal line lies above and to the right of the rectangle, below the power limit, and does not touch other limits. Fig. \ref{trigger_photo} shows an example photo of the trigger circuit we created.

\section{Creating a Charge/Discharge Sequence Circuit with Raspberry Pi and Python}
We created a charging sequence circuit using Raspberry Pi and Python.

\begin{figure*}
\begin{center}
\includegraphics[width=0.8\textwidth]{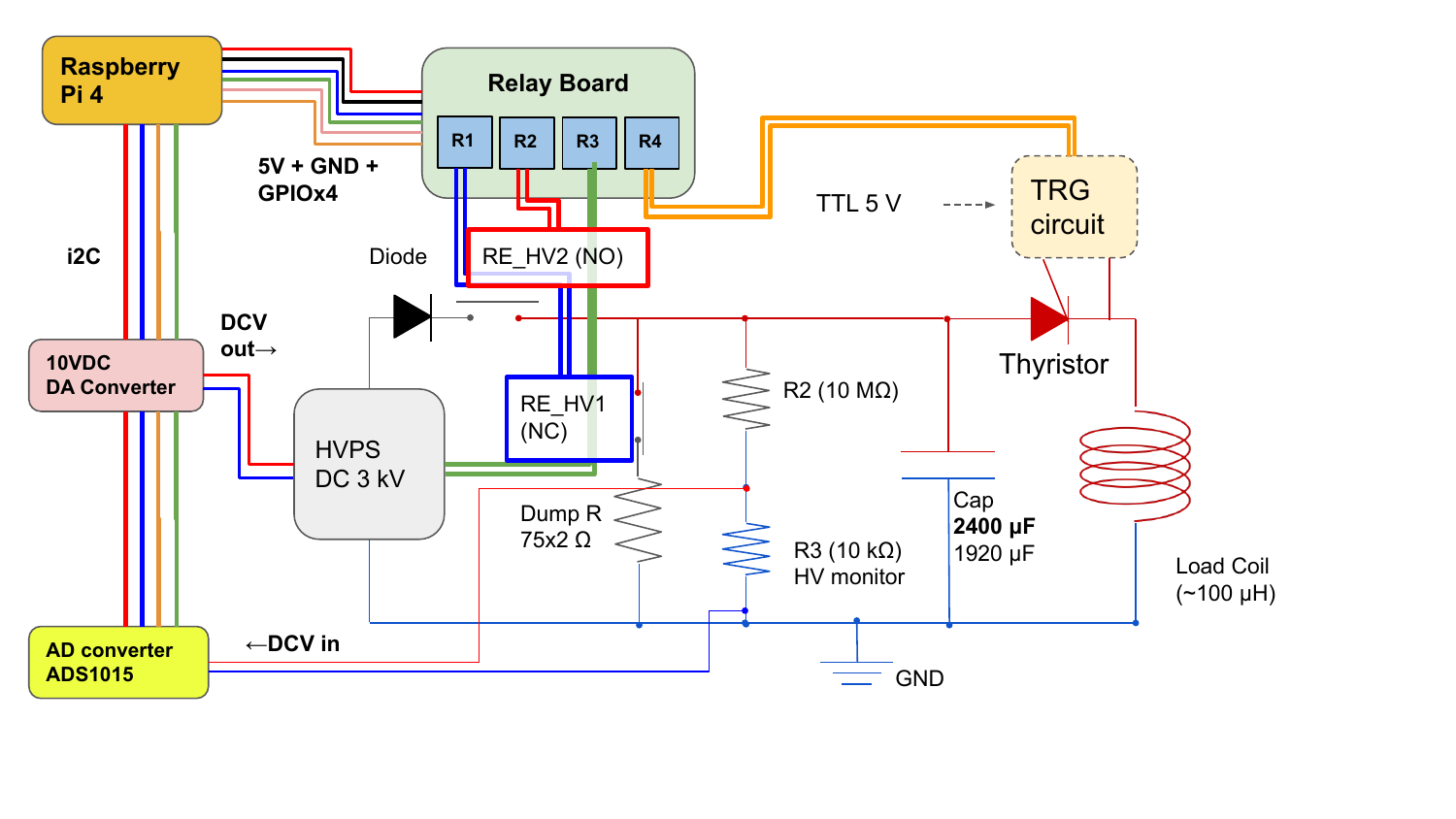}
\caption{Schematic diagram of control system and main circuit\label{controlsystem}}
\end{center}
\end{figure*}

\subsection{The Control Circuit}
Fig. \ref{controlsystem} shows the connection diagram of the control devices and the main circuit. Although the operation sequence of the main circuit has already been shown, more branching logic is needed when considering actual circuit control. For example, when stopping charging, it is necessary to determine whether the capacitor voltage has reached the set voltage. Let's write the charging/discharging sequence in detail again. Refer to this detailed sequence and Fig. \ref{controlsystem} together to understand the circuit's operation.

\begin{enumerate}\label{sequence_detail}
\item[]\underline{Detailed Sequence of Charging and Discharging} (Read with Fig. \ref{controlsystem})
\item Start the control software (GUI) on the Raspberry Pi
    \begin{enumerate}
    \item Initialize the analog voltage for setting the output voltage to the charger to zero volts and output it to the DA converter (DAC)
    \item Initialize the output to the control relays (R1, R2, R3, R4)
    \end{enumerate}
\item Set the charging voltage
    \begin{enumerate}
    \item Convert the charging voltage set in the software to an analog voltage of $0\sim10$ V and output it to the DAC
    \item Display the output charging voltage on the GUI
    \end{enumerate}
\item Start charging
    \begin{enumerate}
    \item When the start charging button is pressed on the GUI, begin the charging operation
    \item Turn on control relay (R1) to open RE-HV1 (Dump relay)
    \item Turn on control relay (R2) to close RE-HV2 (HVPS relay)
    \item Turn on control relay (R3) to start the HVPS (charger) operation
    \item Read the capacitor voltage with an AD converter, convert it to a high voltage value, and display it on the GUI (this is done continuously after the program starts)
    \item When the capacitor voltage reaches 95 \% of the charging voltage, wait 1 second, then turn off control relay (R2) to open RE-HV2 (HVPS relay)
    \item Turn off control relay (R3) to stop the charger operation
    \item Turn on control relay (R4) to charge the trigger circuit and put it in trigger standby mode
    \end{enumerate}
\item Supply an external TTL trigger to the trigger circuit and discharge
\item Immediately after discharge, close RE-HV1 to return to the initial state
    \begin{enumerate}
    \item After charging is complete, monitor the capacitor voltage. When it drops below 5 \% of the charging voltage, determine that discharge has occurred.
    \item Once discharge is confirmed, initialize all control relays.
    \item Display the discharge and initialization status on the GUI.
    \end{enumerate}
\end{enumerate}
Following the sequence list \label{sequence_detail} and Fig. \ref{controlsystem}, you will understand the deeper layer of the main circuit's operation. No further detailed explanation of the sequence will be provided.

Such a circuit is called a sequence circuit, and you frequently encounter them daily (lifts, light switches, etc.). When using it as a device, conditional branching occurs. Therefore, relays are used. High-voltage device relays that interrupt/connect 1000 volts operate with 100 V AC (100 VAC) input. When controlling this with switches or computers, first use relays that can interrupt/connect 100 VAC. These relays operate with a few volts of DC voltage. This low voltage, called weak current, can be controlled by computers or switches. All connections to the control relays in this project use the Normally Open (NO) side. When the relay is on, the circuit is closed, and power is supplied to the high-voltage relay, making it operational.

\subsubsection{How to Implement a Sequence Circuit}
There are three methods to build a sequence circuit:
\begin{enumerate}
\item Design a circuit with many buttons and relays and implement it in a box
\item Use a product called a PLC (Programmable Logic Controller) and program the circuit
\item Use a mini-computer like a Raspberry Pi with a relay array and programming languages such as Python
\end{enumerate}
This time, I wanted to try the third method, so I did. A Raspberry Pi is a small computer embedded on a board, running Linux as the OS. The Raspberry Pi 4 used this time is quite advanced. Initially developed for education and hobbies, half of the current deliveries are to the industrial sector, indicating its widespread use in the industry. Its low price and multifunctionality make it suitable for prototyping new devices. I was initially worried about whether the Raspberry Pi could withstand noise, but it was unfounded. The program, including the GUI, was entirely created in Python. The advantage is the easy availability of information on the Internet. However, the downside is the low support. The second option, PLCs, is the most commonly used worldwide. PLCs are used in small-scale factory automation. Being industrial-grade, they are highly reliable and meet most needs. They also have high support. Although this option was available, we chose the more affordable and beginner-friendly Raspberry Pi. The first method is traditional but challenging to copy or change the logic.

\subsection{Reading Voltage}
\begin{figure}
\begin{center}
\includegraphics[width=0.8\columnwidth]{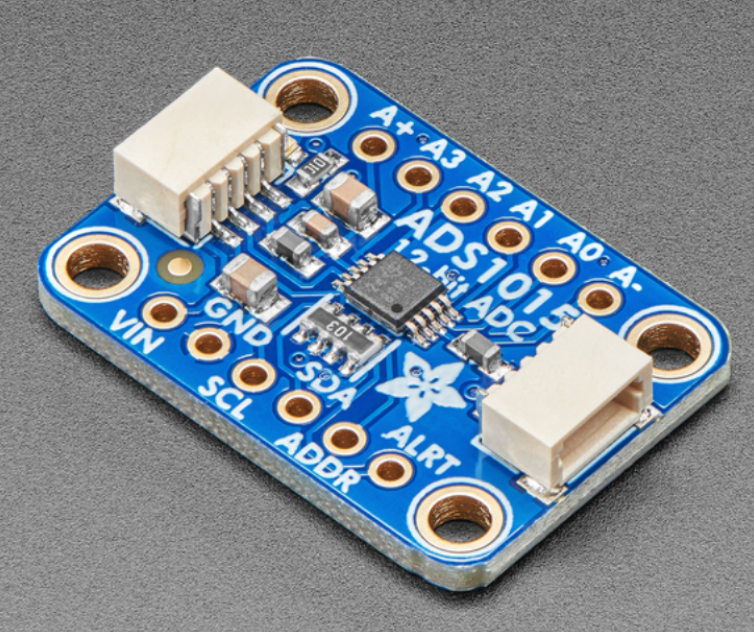}
\caption{ADS1015\label{ADS1015}}
\end{center}
\end{figure}
As shown in Fig. \ref{maincircuit}, the capacitor voltage is obtained by reading a few volts from the voltage divider resistors and multiplying by a constant factor. Since the Raspberry Pi lacks analog input/output, we use a 12-bit 4-channel AD converter called ADS1015\footnote{ADS1115 has 16-bit resolution, so it might have been better}. This device can be controlled via I2C communication, making programming easy. I2C devices can be chain-connected, so the wiring to the Raspberry Pi does not increase. The four terminals used for I2C connection are Vin, GND, SCL, and SDA. This time, we used I2C-compatible AD and DA converters. We also used differential input by utilizing two channels as one. Here is the Python code used.
\begin{lstlisting}[language=Python, caption=adc.py]
import time
import Adafruit_ADS1x15

# ADS1015
def readadc():
    ads = Adafruit_ADS1x15.ADS1015()
    GAIN = 2
    RANGE = 2.048 * 2
    UNIT = RANGE / 4096
    data, volt = 0, 0
    data = ads.read_adc_difference(0, gain = GAIN)
    volt = data * UNIT
    return volt


\end{lstlisting}
It's very short. Note that the library needs to be installed, so in the Raspberry Pi 4 Terminal, run:
\begin{lstlisting}[language=Python, caption=Installation]
pip3 install adafruit-ads1x15
\end{lstlisting}
\footnote{As of October 2023, with the latest Raspberry Pi OS, it is necessary to use the option `--break-system-packages` for installation due to changes in system security.}.
\subsection{Outputting Voltage}
\begin{figure}
\begin{center}
\includegraphics[width=0.8\columnwidth]{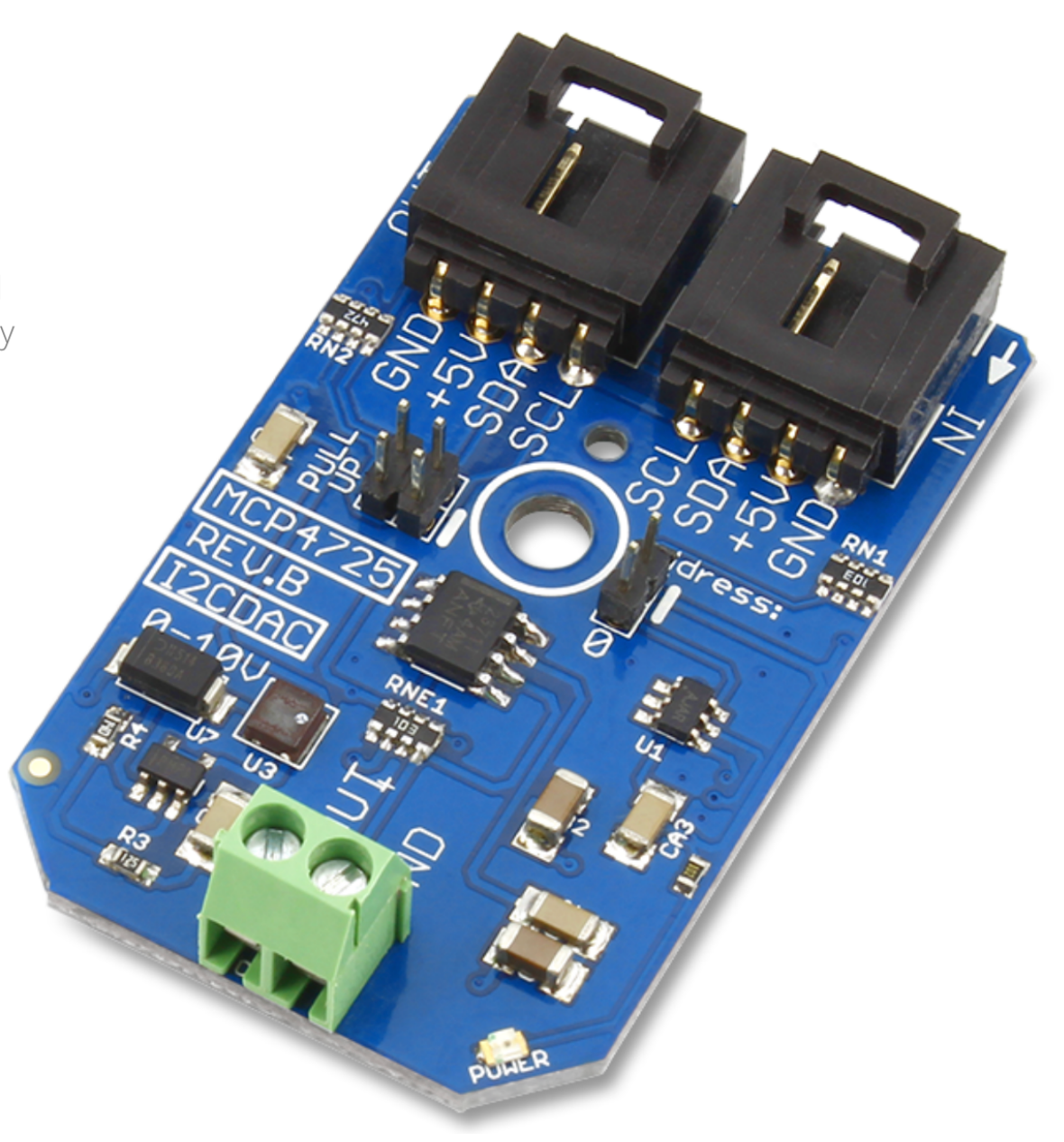}
\caption{Raspberry Pi 0-10V Analog Output Board\label{ADS1015}}
\end{center}
\end{figure}
Controlling and outputting up to 10V DC is not as easy as it seems, but I found a device from National Control Devices called the Raspberry Pi 0-10V Analog Output Board, which I used. This device is convenient because it controls and outputs up to 10V DC via I2C. Here is the Python code to operate it.
\begin{lstlisting}[language=Python, caption=dac10v.py]
#DAC_DC0-10V_TRG
import smbus
import time

def sethvps(hvps):
    
    # Get I2C bus
    bus = smbus.SMBus(1)
    # 0x60 --- i2c address
    # 0x41 -- DAC cahnnel one

    print('HVPS: input voltage from 0 - 3000 V')
    inp0 = float(hvps) # this is to convert 10 v to 3000 v
    inp = inp0 / 300 # this is to convert 10 v to 3000 v

    if inp < 0 or inp > 10:
        print("out of range !!")

    else:
        i = int(inp / 10 * 4095)
        data1 = (i>>4)
        data2 = (i&15)<<4
        data = [data1, data2]
        bus.write_i2c_block_data(0x60, 0x41, data)
        print(f'{inp0} V is written for HVPS.')
\end{lstlisting}
It can be used with standard libraries. The device has an address change function, and by changing 0x60 to 0x61 in the code, two devices can be controlled simultaneously. This program receives a number up to 3000 as the set charging voltage, divides it by 300, and outputs it as a DC voltage up to 10V. The code is short.

\subsection{Relay Control}
\begin{figure}
\begin{center}
\includegraphics[width=0.8\columnwidth]{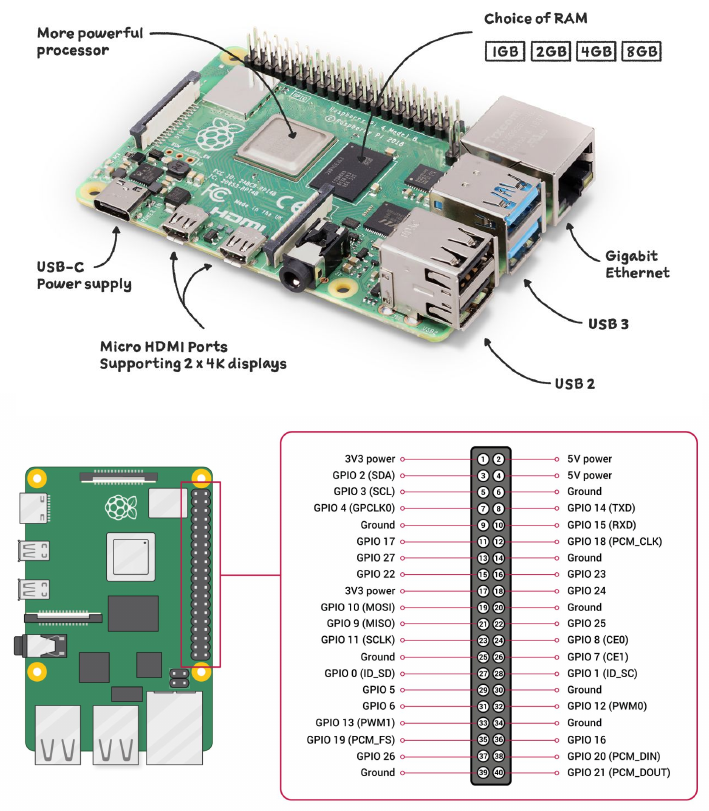}
\caption{Raspberry Pi 4\label{ADS1015}}
\end{center}
\end{figure}
Relays are controlled using the GPIO (General Purpose Input/Output) terminals of the Raspberry Pi. GPIO can output and input Low (0V) and High (3.3V) voltages. Here is an example of Python code to output Low to GPIO.
\begin{lstlisting}[language=Python, caption=gpio.py]
import RPi.GPIO as GPIO

GPIO.setmode(GPIO.BCM)
GPIO.output(4, 0)
GPIO.cleanup(4)
\end{lstlisting}
It's this short. To output High, change 0 to 1. The library is available by default. We used the BitTradeOne Raspberry Pi Relay Control Expansion Board

\subsection{Parts List}
Readers can obtain the complete list of the parts used upon request to the corresponding author.


\section{Mini Coil Fabrication}
\subsection{Fabrication Steps}
Once the capacitor bank is completed, let's make the coil. Before starting the mini coil fabrication, decide on the following:
\begin{itemize}
\item The maximum intended magnetic field (in tesla)
\item Inner diameter (bore diameter)
\item Wire diameter
\item Number of turns (in the axial direction)
\item Number of layers
\end{itemize}
For a maximum of 20 T, copper wire is sufficient. For the wire coating, it is recommended to use AIW coating\footnote{Polyimide-amide enameled wire}, which has high electrical insulation and mechanical wear properties. Copper wire with a 1.0 mm diameter and AIW coating is inexpensive and readily available, making it a good starting point. Copper wire can generate magnetic fields up to about 20 T. For fields above 20 T, adequate reinforcement is necessary to prevent coil deformation.

For coils that generate up to 40 T, choose a copper-silver alloy wire (10 \% silver). The coating should also be AIW. This time, a 1.0 mm diameter round wire was used. This wire is custom-made and expensive, so consider joint purchasing or consulting researchers who already have it. Even with copper-silver alloy wire, generating fields above 20 T will result in significant coil heating. Therefore, the entire coil needs to be immersed in liquid nitrogen. Liquid nitrogen has high insulation properties, and it is desirable to immerse the electrodes near the coil as well.

The coil requires flanges to secure the electrodes. While FRP is widely used for flange material, PLA filament for 3D printers has also proven quite useful\footnote{It holds up reasonably well at low temperatures but may crack after repeated use in low-temperature environments. However, it is effective for prototyping.}. For the core rod, use a stainless steel rod sprayed with Teflon. Fix the flanges to this rod in advance using Stycast 1266.

\subsection{Simple Semi-Automatic Winder}
While coils can be wound by hand, a winding machine is very useful. A winding mechanism to wind the wire onto the coil is sufficient, but a rewinding mechanism is also convenient. A winding mechanism ensures constant tension on the wire and allows rewinding if necessary. Detailed information is available on Oriental Motor's website. By combining a speed control motor and a torque control motor, tension can be applied while winding the coil. Here is a view of the semi-automatic winder created. The speed control motor is operated with a foot pedal. This setup frees both hands during winding, making it very convenient for applying adhesive or fine-tuning the wire position. Copper-silver alloy wire is harder than copper wire and requires sufficient tension that cannot be easily applied by hand.
\begin{figure}
\begin{center}
\includegraphics[width=\columnwidth]{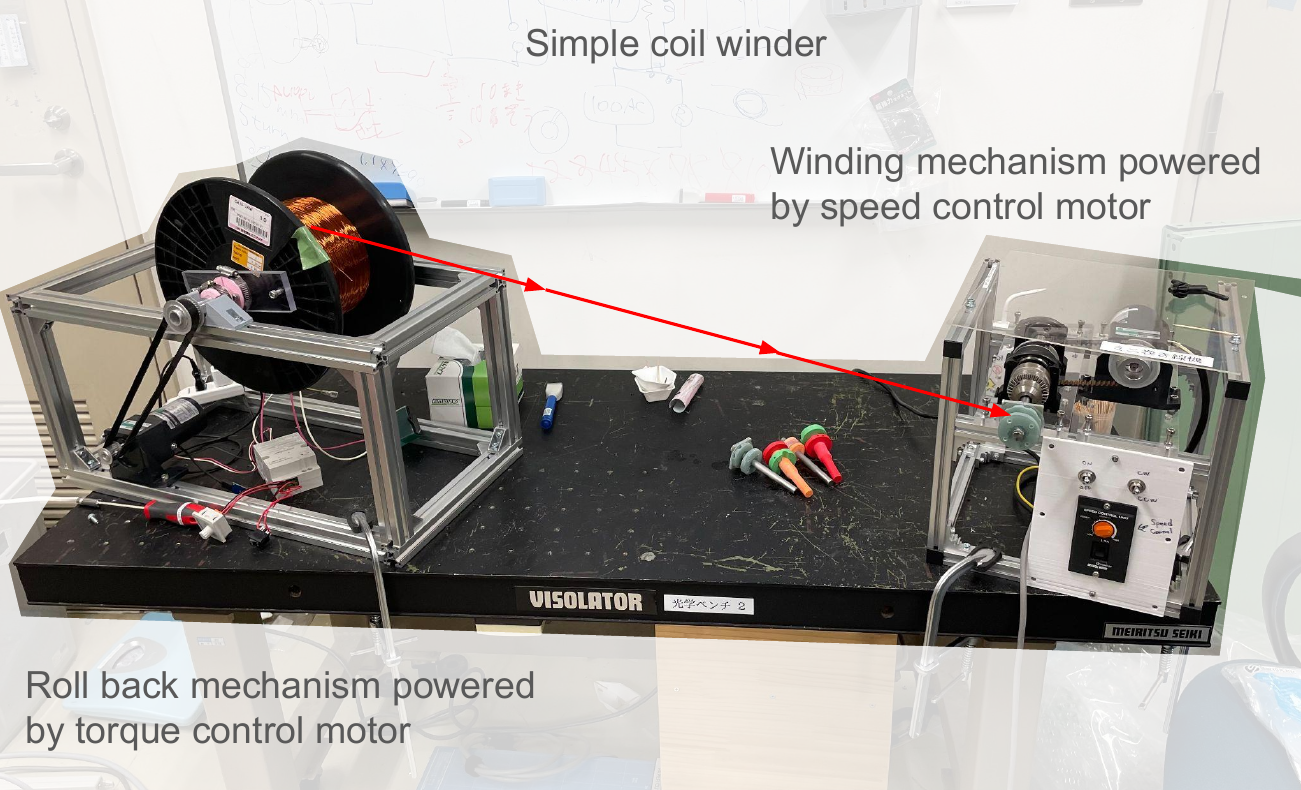}
\caption{Photo of a home-made winder using a speed control motor and a torque control motor.\label{winder}}
\end{center}
\end{figure}

\begin{figure}
\begin{center}
\includegraphics[width=\columnwidth]{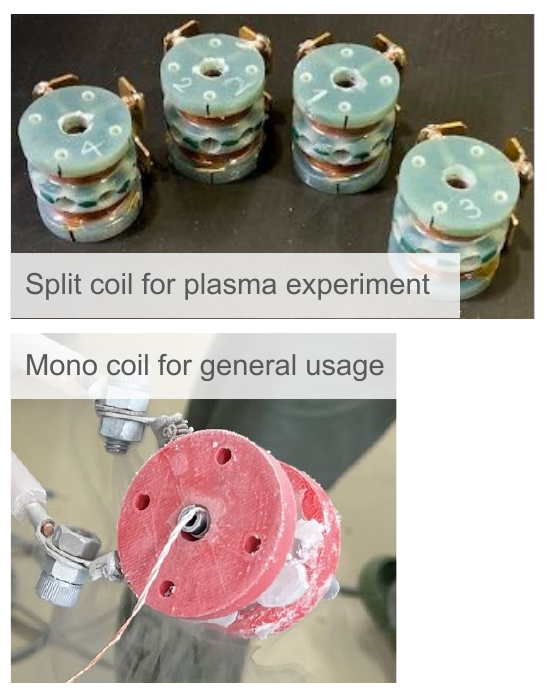}
\caption{Photo of handmade coils\label{winder}}
\end{center}
\end{figure}

\subsection{Evaluation of Mini Bank and Coil}
First, perform a charging and grounding test for the mini bank itself. Conduct this test in a spacious area with multiple people. Gradually increase the charging voltage, grounding each time. Charge up to the target voltage and wait for about three times longer than the usual charging time before grounding. Ensure that you can safely ground the electricity at any time without discharging it into the coil.

Next, test the coil. Begin by testing the coil with low voltage and low current. At this stage, verify that the pulse width, current, and generated magnetic field roughly match the calculated values. If there are significant discrepancies, something might be wrong with the coil choice, wiring, or measurement system. Gradually increase the charging voltage. Initially, plot the charging voltage against the maximum current and magnetic field, confirming a linear relationship. As you approach the maximum voltage, observe waveform distortions and loss of linearity in the previous plot due to coil heating. Estimate the maximum coil temperature from measurements or calculations, and if it exceeds room temperature by 50 K, it might be time to stop. Carefully observe for abnormal noise in the current and magnetic field signals and note any changes over time. Abnormalities may be caused by voltage breakdown or wire movement due to magnetic forces. The voltage between coil terminals is highest at $ t = 0 $ and zero at the time when the current reaches its peak $ t = t_{\rm peak} $. Conversely, the force on the coil is zero at $ t = 0 $ and maximizes at $ t = t_{\rm peak} $. If the coil gets damaged, consider whether the issue occurred around $ t = 0 $ or $ t = t_{\rm peak} $ to determine if the cause was voltage, force, or possibly heat. After testing, inspect the coil visually and check for any damage.

\section{Measurement of Current, Magnetic Field, and Physical Properties}	
\subsection{Measurement of Pulsed Current}
\begin{figure}
\begin{center}
\includegraphics[width=0.8\columnwidth]{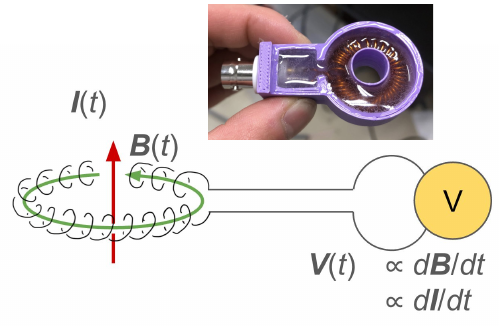}
\caption{Schematic diagram and photo of a homemade pickup coil for pulsed current measurement\label{meas_i}}
\end{center}
\end{figure}
For measuring pulsed current, you can use commercial current transformers or Rogowski coils, as well as homemade Rogowski coils\footnote{Products from Pearson or Power Electronic Measurements are popular.}. All of these methods involve wrapping a coil around the current to pick up the magnetic field generated around it and measuring the induced voltage. The induced voltage is proportional to the time variation of the magnetic flux, resulting in a waveform that is the derivative of the current. Commercial products also include an analog integrator that outputs the current waveform directly. If you make one yourself, simply wrap insulated wire around a rod, slip it off, and wrap it around the wire. A voltage will appear at both ends of the wire, which can be measured. It's good to have a commercial product for calibration at low currents. When making your own, be sure to pay attention to insulation. An example of a homemade current measuring coil is shown.

\subsection{Measurement of Pulsed Magnetic Field}	
\begin{figure}
\begin{center}
\includegraphics[width=0.8\columnwidth]{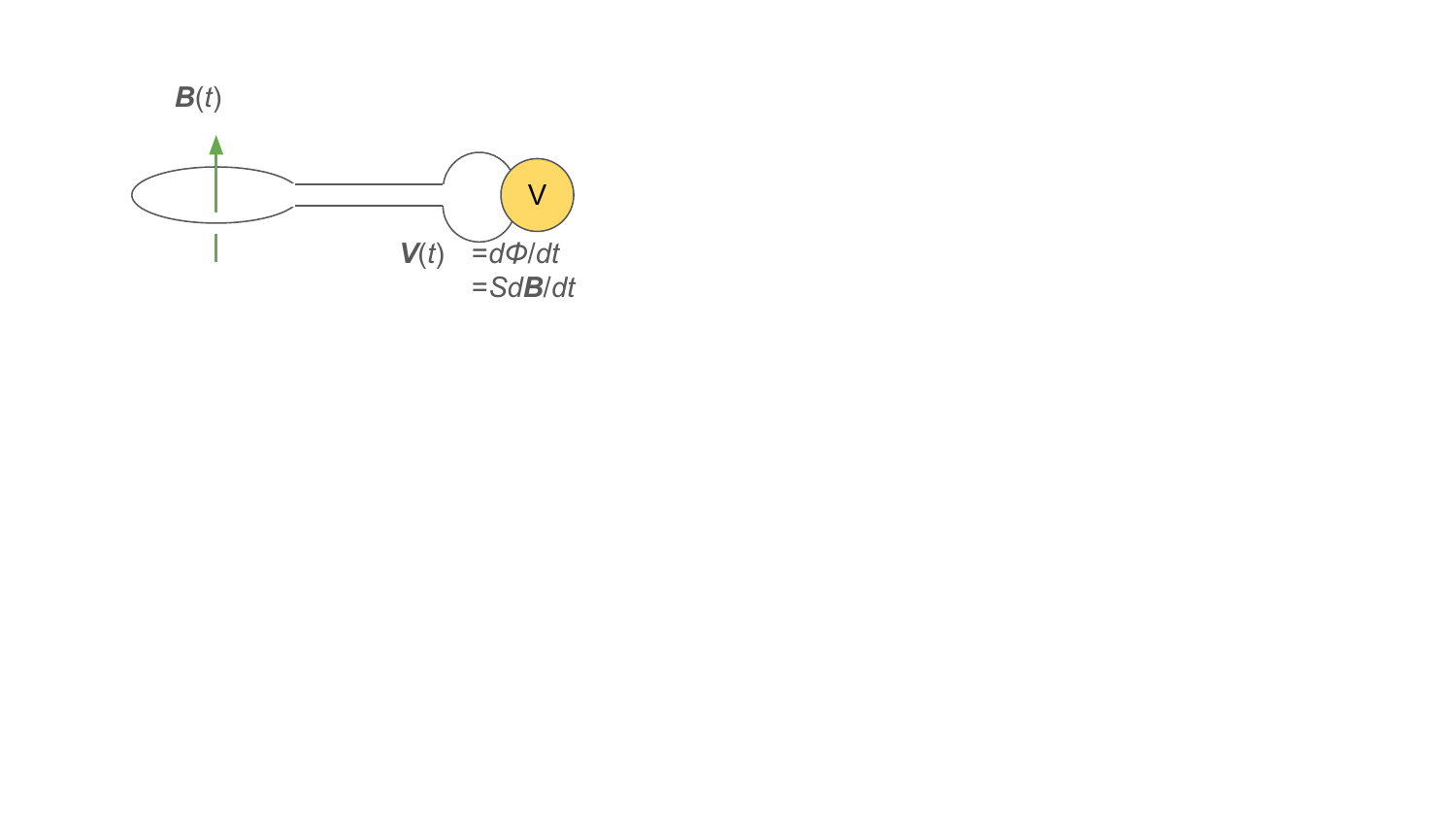}
\caption{Schematic diagram and photo of a pickup coil for pulsed magnetic field measurement\label{meas_b}}
\end{center}
\end{figure}
\begin{figure*}
\begin{center}
\includegraphics[width=\textwidth]{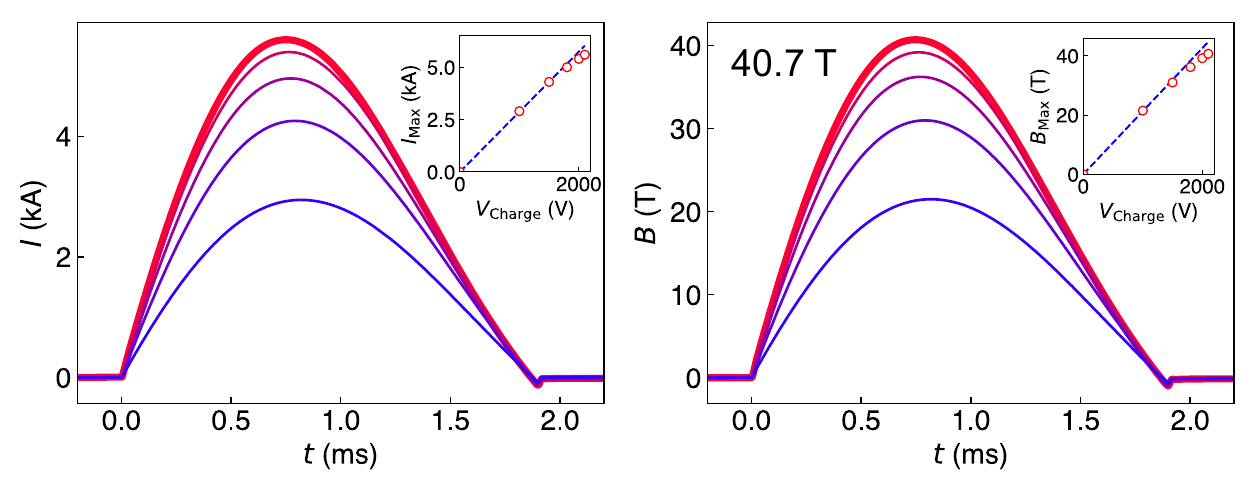}
\caption{Measurement results of current and magnetic field generated by the mini capacitor bank and mini coil produced in this study.\label{ibplot}}
\end{center}
\end{figure*}

When a pulsed current is input to a coil, a pulsed magnetic field is generated. Since a pulsed magnetic field varies with time, it can also be easily measured with a pickup coil. Place a coil inside the main coil so that it is penetrated by the magnetic field.
\begin{equation}
V = \frac{d\Phi}{dt} = \frac{d(BS)}{dt}
\end{equation}
The induced voltage in the pickup coil is the time variation of the total magnetic flux $\Phi$. The relationship $\Phi = B S$ holds between the magnetic flux density $B$ and the effective cross-sectional area $S = s \times n$ (the product of the pickup coil's cross-sectional area $s$ and the number of turns $n$). To calculate the magnetic field from the voltage measured with an oscilloscope:
\begin{equation}
B(t) = \int^{t}_{t = 0}\frac{V(t)}{S}
\end{equation}
and perform numerical integration. An example of the measured current and magnetic field is shown in Fig. \ref{ibplot}.

\begin{figure}
\begin{center}
\includegraphics[width=0.8\columnwidth]{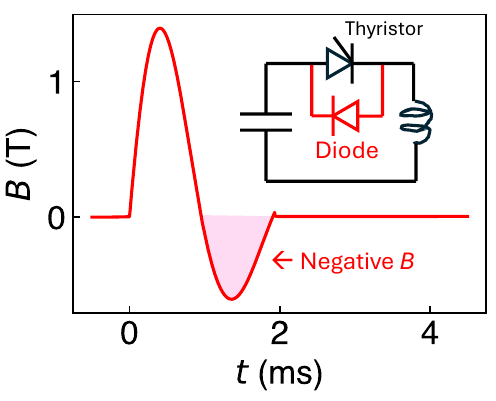}
\caption{Example of a negative pulse magnetic field waveform achieved with a reverse diode\label{inv}}
\end{center}
\end{figure}
The actual discharge was performed with the parameters (4) calculated earlier, matching the parameters of the coil and capacitor bank. Although the calculation predicted the generation of 43.6 T, the actual generation was only 39.2 T. As seen in the inset of Fig. \ref{ibplot}, the maximum current and magnetic field values are capped and deviate from the linear relationship with the charging voltage. This is due to heating, indicating the limit of the current coil. The calculation assumed the electrical resistance of copper, but since copper-silver wire has a higher resistance than copper, the heating was underestimated. By increasing the charging voltage from 2000 V to 2100 V, we confirmed the generation of 40.7 T. Thus, a 40 T device was successfully completed.

Finally, data showing a magnetic field waveform generated over one cycle, including a negative magnetic field, is presented (Fig. \ref{inv}). The thyristor has a rectifying effect, usually generating a half-cycle pulse. By placing a diode in parallel and in the opposite direction to the thyristor, reverse current can flow, producing a full-cycle pulse. This allows the generation of negative magnetic fields without changing the setup, making it useful for studying the hysteresis of ferromagnetic materials near zero magnetic field or for research on multiferroics that require reverse magnetic field data.

\subsection{Measurement of Physical Properties under Pulsed Magnetic Fields}	
\subsubsection{Magnetization}

\begin{figure}
\begin{center}
\includegraphics[width = 0.8\columnwidth]{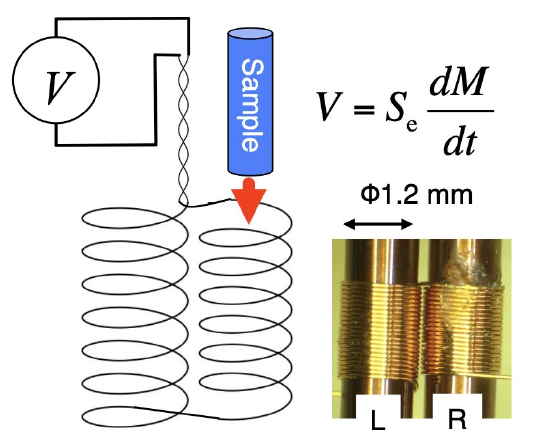}
\caption{A schematic drawing of magnetization measurement \label{meas_m}}
\end{center}
\end{figure}

Under a pulsed magnetic field, the magnetization of the sample also changes with time, generating an induced electromotive force in the pickup coil. Two pairs of coil pickups are connected in series to cancel the external magnetic field. It can be arranged side by side or coaxially. Since a complete pickup coil cannot be made, perform a pulse measurement once without the sample, and then perform a measurement with the sample, and subtract the data. The obtained voltage has information proportional to $\frac{dM}{dt}$. By integrating this, $M(t) \propto \int^{t}_{t=0} V(t) dt$ is obtained. By using the data of $M(t)$ and $B(t)$ obtained, $M(B)$ can be plotted. This is the magnetization curve under a strong magnetic field. Since this can also be converted to $\mu(B)$ using $B = \mu H$, this data can be used as a basis for identifying magnetic materials.

\subsubsection{Electrical Resistivity}

\begin{figure}
\begin{center}
\includegraphics[width = 0.8\columnwidth]{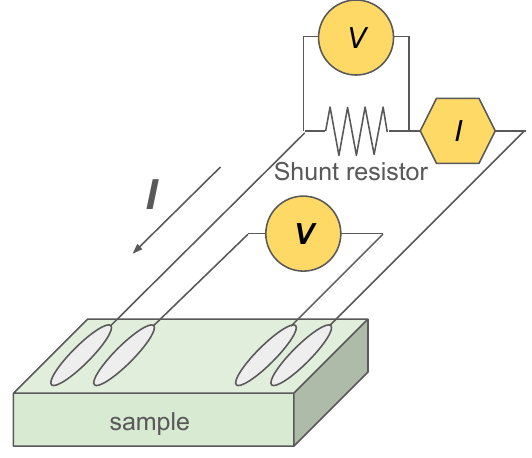}
\caption{A schematic drawing of magnetoresistance measurement \label{meas_r}}
\end{center}
\end{figure}

When the electrical resistance of the sample ranges from a few ohms to several hundred ohms, it is measured using the four-terminal method. Voltage terminals are attached to the sample to measure the voltage. Current terminals are also attached. To measure the current, a shunt resistor is placed near the power source. The current is calculated by measuring the voltage across the shunt resistor. This allows the voltage and current of the sample to be known, thus enabling the measurement of electrical resistance. Similar to magnetization measurements, loops can form around the wires near the sample, inevitably leading to a background magnetic field. To avoid this, AC measurements can be performed using an alternating current source \footnote{A signal generator is sufficient}. If the frequency is sufficiently high, 100 kHz or above, relative to millisecond magnetic field pulses, unnecessary background and noise can be eliminated after data acquisition by performing numerical lock-in detection, allowing for single-shot measurements.

\subsubsection{Magnetostriction}

\begin{figure}
\begin{center}
\includegraphics[width = 0.8\columnwidth]{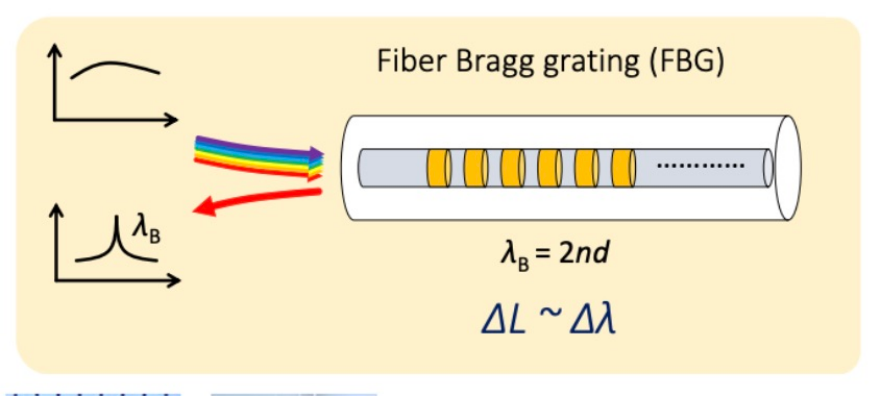}
\caption{A schematic drawing of a FBG. The elongation and the shrinkage of the optical fiber can be detected by the change of the Bragg wavelength $\lambda_{\rm{B}}$.
By gluing a sample to the fiber, the strain of the sample is monitored optically from a distance. \label{meas_fbg}}
\end{center}
\end{figure}

Traditionally, resistance measurement using strain gauges and capacitance dilatometory have been commonly used for magnetostriction measurements.
Recently, a method for measuring magnetostriction using optical fibers with strain measurement capability, called Fiber Bragg Grating (FBG), has become available.
Preparation around the sample involves simply attaching the FBG fiber to the sample with adhesive and placing it in a magnetic field.
A wideband light source is coupled into the FBG fiber (e.g., ASE light source or SLD light source).
The FBG reflects monochromatic light at $\lambda_{\mathrm{B}}$ (Bragg reflection).
The change in $\lambda_{\mathrm{B}}$ contains information about the strain of the sample, $\Delta\lambda_{\mathrm{B}} \propto \Delta L/L$.
To measure this, there are methods using a high-speed infrared spectrometer and a point detector.
In the latter case, in order to relate the obtained light intensity information $I_{\mathrm{opt}}$ to $\Delta\lambda_{\mathrm{B}}$, the spectrum of the wideband light source is shaped using the edge of an optical filter.
This creates a situation where $\Delta I_{\mathrm{opt}} \propto \Delta\lambda_{\mathrm{B}} \propto \Delta L/L$.
By measuring $I_{\mathrm{opt}}$, strain measurement becomes possible.

\section{Low-Temperature Experiments}
An example of a small-scale cryogenic device is introduced, which utilizes a plastic He flow cryostat.
The tip of a small vacuum container is inserted into a coil.
Inside the vacuum container, a plastic He flow cryostat is installed.
The plastic He flow cryostat is designed to allow top-loading of probes, enabling various physical property measurements in low-temperature and magnetic field environments.
The vacuum chamber was 3D printed and coated internally with Nitofix \footnote{Nitofix SK-229 (Nitto Shinko)}.
The cryostat was constructed entirely of FRP Fiber reinforced plastic (FRP), also bonded with Nitofix.

Considering recent helium supply shortages, price increases, and difficulty in obtaining helium as a coolant, experiments using a GM refrigerator may be the most suitable option, especially for small-scale experiments.
The role of the GM refrigerator is to cool the sample, and if the coil cooling is considered sufficient with liquid nitrogen, then it seems feasible somehow.

\section{Conclusion}
As described above, it is possible to easily and cost-effectively produce a lightweight 40-tesla pulsed magnetic field device weighing about 50 kg. This mini bank can be used not only for rough physical property measurements up to 40 T. Also, it can be dedicated to truly challenging and new experiments such as applying magnetic fields to plasma, vacuum, chemical reactions, and life sciences. It can also be used for experiments utilizing multiple capacitor banks simultaneously. Our group is planning applications beyond property evaluation in high magnetic fields, such as plasma experiments and life science experiments, and we are also prototyping a vector pulse magnet driven by two mini banks. The possibilities opened up by mini banks are vast. We wish our readers a happy high magnetic field life.
We hope young researchers and beginners in pulsed magnetic fields will master this technology and pioneer new experimental areas in high magnetic fields. 
\bibliography{mini}
\appendix
\section{Energy of magnetic field}	
\begin{lstlisting}[language = Python, label = mage, caption = magneticfieldenergy.py]
import matplotlib.pyplot as plt
import numpy as np

config = {
    'lines.linewidth':1.5,
    'xtick.direction': 'in',
    'ytick.direction': 'in',
    'font.sans-serif': ['Helvetica', 'Arial'],
    'figure.dpi': 200,
}
plt.rcParams.update(config)

mu0 = 4 * np.pi * 1e-7

def energy(B, V):
    return B ** 2 * V / mu0 / 2

fig, ax = plt.subplots(1, 1
                       , tight_layout = 1
                       , figsize = (3.6, 3.2)
                      )

B = np.arange(0, 101, 1)
V = 1e-6 * 0.5 * 0.5 * 1.0 * np.pi
ax.plot(B, energy(B, V) / 1000, color ='r')
ax.set_ylabel('$E$ (kJ)')
ax.set_xlabel('$B$ (T)')
ax.annotate('Cylinder of $\phi$ 1.0 cm, L = 1.0 cm\n $E = B^2V/(2 \mu_0)$', (10, 2.7))
ax.set_xlim(0, 100)
ax.set_ylim(0, 3.5)
ax.vlines(40, 0, 0.5, ls = '--', lw = 1)
ax.hlines(0.5, 0, 40, ls = '--', lw = 1)
ax.vlines(80, 0, 2, ls = '--', lw = 1)
ax.hlines(2, 0, 80, ls = '--', lw = 1)


# plt.suptitle('energy stored in a magnetic field')
plt.savefig('energy_B_single.pdf')
plt.savefig('energy_B_single.png')
\end{lstlisting}
\section{LCR simulation using Python}	
\begin{lstlisting}[language = Python, label = simulation, caption = pulsemagnet\_simulation.py]
# Import necessary libraries
import numpy as np
import matplotlib.pyplot as plt

# Set up Matplotlib configurations for consistent plot styling
config = {
    'lines.linewidth': 1.5,
    'xtick.direction': 'in',
    'ytick.direction': 'in',
    'font.sans-serif': ['Helvetica', 'Arial'],
    'figure.dpi': 200,
}
plt.rcParams.update(config)

# Define physical constants and input parameters for the simulation
wire = 1  # Wire diameter in mm
wire_cover = 0.033  # Insulation thickness in mm
n = 16  # Number of windings per layer
m = 10  # Number of layers
bore = 10  # Inner diameter of the coil in mm
resty0 = 5e-10  # Base resistivity in ohm m
resty1 = 6.5e-11  # Resistivity coefficient per T in ohm m/T
T_D = 343  # Debye temperature of copper in K
cap_single = 800e-6  # Capacitance of a single capacitor in F
cap_num = 3  # Number of capacitors
cap = cap_single * cap_num  # Total capacitance in F
v0 = 2000  # Initial voltage in V
q0 = v0 * cap  # Initial charge in C
tp0 = 78  # Initial temperature in Kelvin
t0 = 0.0  # Start time of the simulation in s
t1 = 20e-3  # End time of the simulation in s
del_t = 5e-6  # Time step for the simulation in s

# Additional geometric and physical parameters
F_t = 0  # Thickness of the flange (not used in calculations)
z = 0  # Measurement position along the z-axis in mm
R_r = 0  # Residual resistance in mOhm
L_r = 2.5  # Residual inductance in uH

# Calculate derived geometrical parameters
r1 = (bore * 1e-3) / 2  # Inner radius in meters
r2 = m * (wire + wire_cover * 2) * 1e-3 + r1  # Outer radius in meters
c_r = r2 * 2 * 1e3  # Outer diameter in mm
l1 = (wire + wire_cover * 2) * 1e-3 * n / 2  # Half-height of the coil in meters
coil_length = l1 * 2 * 1e3  # Total height of the coil in mm
alpha = r2 / r1  # Ratio of outer to inner radii
beta = l1 / r1  # Aspect ratio of the coil
mu0 = 4 * np.pi * 1e-7  # Magnetic constant (permeability of free space)
K = (1 + 0.225 * (alpha + 1) / beta + 0.64 * (alpha - 1) / (alpha + 1) + 0.42 * (alpha - 1) / beta) ** -1  # Nagaoka's constant
L = mu0 * np.pi / 8 * r1 * (alpha + 1) ** 2 / beta * (n * m) ** 2 * K + L_r * 1e-6  # Total inductance in Henries

# Calculate the cross-sectional area and length of wire
S = (wire * 1e-3 / 2) ** 2 * np.pi  # Cross-sectional area in m^2
l_length = 0  # Total length of the wire in m
for m0 in np.arange(m):
    l_length += 2 * np.pi * n * (r1 + (m0 + 1 / 2) * (wire + wire_cover * 2) * 1e-3)  # Sum up each layer's length

# Calculate the position and total moles of copper in the wire
z0 = z * 1e-3  # Position of the B-sensor from the field center in meters
R_const = 8.31447  # Universal gas constant in J/K mol
density = 8920 * 1e3  # Density of copper in g/m^3
mol = density * l_length * S / 63.536  # Total moles of copper

# Define simulation functions
def resty(tp):
    """Calculate resistivity at temperature tp."""
    return resty0 + tp * resty1

def r(tp):
    """Calculate resistance at temperature tp."""
    return resty(tp) * l_length / S

def f(q, i, tp, t):
    """Calculate the derivative of charge with respect to time."""
    return -q / cap - r(tp) * i

def debye(x):
    """Debye model for specific heat."""
    return x ** 4 * np.exp(x) / (np.exp(x) - 1) ** 2

# Set initial conditions and run simulation
qpoints, ipoints, tppoints, tpoints = np.array([]), np.array([]), np.array([]), np.arange(t0, t1, del_t)
i0 = 0.0  # Initial current in A
q, i, tp = q0, i0, tp0

# Runge-Kutta integration for system of differential equations
for t in tpoints:
    # Implement the 4th-order Runge-Kutta method for numerical integration
    k1i = f(q, i, tp, t) * del_t / L
    k1q = i * del_t
    k2i = f(q + k1q / 2, i + k1i / 2, tp, t + del_t / 2) * del_t / L
    k2q = (i + k1i / 2) * del_t
    k3i = f(q + k2q / 2, i + k2i / 2, tp, t + del_t / 2) * del_t / L
    k3q = (i + k2i / 2) * del_t
    k4i = f(q + k3q, i + k3i, tp, t + del_t) * del_t / L
    k4q = (i + k3i) * del_t

    tp += r(tp) * i ** 2 * del_t / hcap(tp)  # Update temperature
    i += (k1i + 2 * k2i + 2 * k3i + k4i) / 6  # Update current
    q += (k1q + 2 * k2q + 2 * k3q + k4q) / 6  # Update charge

# Calculate magnetic field strength using Biot-Savart Law adjustments
ipoints = -1 * ipoints
delta = z0 / r1
KBp = (beta + delta) / (alpha - 1) * np.log((alpha + (alpha ** 2 + (beta + delta) ** 2) ** 0.5) / (1 + (1 + (beta + delta) ** 2) ** 0.5))
KBm = (beta - delta) / (alpha - 1) * np.log((alpha + (alpha ** 2 + (beta - delta) ** 2) ** 0.5) / (1 + (1 + (beta - delta) ** 2) ** 0.5))
bf = mu0 * n * m * ipoints / (4 * l1) * (KBp + KBm)  # Magnetic field along z-axis

# Plot and save the results
np.savetxt('out.csv', bf)  # Save output data to CSV

# Determine maxima and minima for display in plots
maxx = np.argmax(bf)  # Index of max magnetic field
maxt = round(tpoints[maxx] * 1000, 2)  # Time of max field in ms
plsx = np.argmin(np.abs(bf[maxx * 2: maxx * 3])) + maxx * 2  # Index of pulse width end
plswid = round(tpoints[plsx] * 1000, 2)  # Pulse width in ms

imax = round(ipoints[maxx] / 1000, 1)  # Max current in kA
vmax = v0  # Max voltage (constant)
bmax = round(bf[maxx], 1)  # Max magnetic field in T
imin = round(ipoints[int(plsx * 1.1)] / 1000, 1)  # Min current post-pulse in kA
vmin = round(qpoints[int(plsx * 1.1)] / cap, 1)  # Min voltage post-pulse
bmin = round(bf[int(plsx * 1.1)], 1)  # Min magnetic field post-pulse
tmax = round(tppoints[plsx], 1)  # Max temperature during pulse
tmin = tp0  # Initial temperature

efficiency = round(bmax / imax, 1)  # Efficiency of magnet in T/kA
tpos1 = maxt * 0.5
tpos2 = maxt * 1.5

llabel = [['$B$ (T)', '$V$ (V)'], ['$I$ (kA)', '$T$ (K)']]
maxs = [[bmax, vmax], [imax, tmax]]
mins = [[bmin, vmin], [imin, tmin]]

# Create subplots for each variable
fig, axs = plt.subplots(2, 2, figsize=(8, 6), tight_layout=True)
axs[0, 1].plot(tpoints * 1000, qpoints / cap, '.', color='red')  # Plot voltage over time
axs[0, 1].text(tpos2, vmax * 0.8, f'max: {maxt} ms')
axs[0, 1].text(tpos2, vmax * 0.5, f'pulse: {plswid} ms')
axs[1, 0].plot(tpoints * 1000, ipoints / 1000, '.', color='blue')  # Plot current over time
axs[1, 0].text(tpos1, imax * 0.1, f'I_max: {imax} kA')
axs[0, 0].plot(tpoints * 1000, bf, '.', color='green')  # Plot magnetic field over time
axs[0, 0].text(tpos1, bmax * 0.4, f'B_max: {bmax} T')
axs[0, 0].text(tpos1, bmax * 0.1, f'Efficiency: {efficiency} T/kA')
axs[1, 1].plot(tpoints * 1000, tppoints, '.', color='black')  # Plot temperature over time
axs[1, 1].text(tpos2, tmin * 0.9, f'T_max: {tmax} K')

for i in range(2):
    for j in range(2):
        axs[i, j].set_xlabel('$t$ (ms)', fontsize=12)
        axs[i, j].set_ylabel(llabel[i][j], fontsize=12)
        axs[i, j].set_ylim(mins[i][j] - abs(mins[i][j]) * 0.2, maxs[i][j] * 1.2)
        axs[i, j].set_xlim(-0.1, plswid * 1.1)

fig.suptitle('Pulse magnet simulator')
plt.tight_layout()
plt.savefig('fig.pdf')  # Save figure to file
plt.show()

# Output simulation results
print('---Inputs---')
print(f'coil_height is {round(coil_length, 1)} mm.')
print(f'coil_diameter is {round(c_r, 1)} mm.')
print(f'bore is {round(bore, 1)} mm.')
print(f'L is {round(L * 1e6, 1)} \muH.')
print(f'(L_res is {round(L_r, 1)} \muH.)')
print(f'C is {round(cap * 1e6, 1)} \muF.')
print(f'V_0 is {round(v0, 1)} V.')
print(f'R_res is {round(R_r, 1)} mohm.')

print('\n---Results---')
print(f'Total length of magnet wire is {round(l_length, 1)} m.')
print(f'I_max is {round(np.amax(ipoints))} A.')
print(f'B_max is {round(np.amax(bf), 1)} T.')
print(f'T_ini is {round(tp0, 1)} K.')
print(f'T_end is {round(tp, 1)} K.')
print(f'R_ini is {round(r(tp0) * 1000, 1)} mohm.')
print(f'R_end is {round(r(tp) * 1000, 1)} mohm.')
print(f'C_v  is {round(hcap(tp), 1)} J/K')
print(f'E_tot  is {round(cap * v0 ** 2 / 2)} J')
\end{lstlisting}
\section{GUI controller for Pi-MAG and Pi-TOP}	
\begin{lstlisting}[language = Python, label = simulation, caption = control\_main\_v4.py]
from tkinter import *
from tkinter import ttk
import RPi.GPIO as GPIO 
import time
import dac10v
import adc

# GUI level settings for widget placement
title_level = 0
hvps_level = 2
hvps2_level = 3
vread_level = 4
control_level = 5
auto_control_level = 6
exit_level = 7
logo_level = 8

waitasec = 0.1  # Wait time used in delays

# Initialize High Voltage Power Supply (HVPS) setting
hvps = 0
dac10v.sethvps(hvps)  # Set initial HVPS value

# Setup GPIO for control pins
gpio_relay = [4, 17, 27, 22]  # Define GPIO pins
GPIO.setmode(GPIO.BCM)  # Set GPIO mode to BCM
for i in range(4):
    GPIO.setup(gpio_relay[i], GPIO.OUT)
    GPIO.output(gpio_relay[i], 0)  # Initialize all relays to OFF

# Initialize the main GUI window
root = Tk()
root.title('PiMag Control')  # Title of the window

# Setup main frame for widgets
frame1 = ttk.Frame(root, padding=(30))
frame1.grid()

# Title and logo configuration
img = PhotoImage(file="logo2.png")  # Load an image for the logo
title_label = ttk.Label(frame1, padding=(24), text='Pi-MAG: a Portable 40 T generator', font=('San-Serif', '40', 'bold'))
logo_label = ttk.Label(frame1, image=img)
title_label.grid(row=title_level, column=0, columnspan=4)

# Styling for buttons and checkboxes
style = ttk.Style()
style.configure('TButton', font=('', 24))
style.configure('TCheckbutton', font=('', 24))

# HVPS setting UI elements
hvps_var = StringVar(value='0')
def hvps_set():
    """Set HVPS value based on user input."""
    global hvps
    hvps = hvps_var.get()
    dac10v.sethvps(hvps)
    hvps = str(float(hvps))
    hvps_label1['text'] = hvps
    
# HVPS control buttons for voltage adjustment
def p100(): increment_voltage(100)
def m100(): decrement_voltage(100)
def p10(): increment_voltage(10)
def m10(): decrement_voltage(10)

# HVPS UI setup
hvps_label = ttk.Label(frame1, padding=(20), text='HVPS (0 ~ 2000 V):', font=('default', '24'))
hvps_entry = ttk.Entry(frame1, textvariable=hvps_var, width=6, font=('default', '24'))
hvps_label1 = ttk.Label(frame1, padding=(20), text='0', font=('default', '24'))
hvps_label2 = ttk.Label(frame1, padding=(20), text='V is set.', font=('default', '24'))
set_btn = ttk.Button(frame1, padding=(20), text='Set', command=hvps_set)
hvps_label.grid(row=hvps_level, column=0)
hvps_entry.grid(row=hvps_level, column=1)
set_btn.grid(row=hvps_level, column=2)
hvps_label1.grid(row=hvps_level, column=3)
hvps_label2.grid(row=hvps_level, column=4)

# Manual control buttons for quick voltage changes
p100_btn = ttk.Button(frame1, padding=(20), text='+100', command=p100)
m100_btn = ttk.Button(frame1, padding=(20), text='-100', command=m100)
p10_btn = ttk.Button(frame1, padding=(20), text='+10', command=p10)
m10_btn = ttk.Button(frame1, padding=(20), text='-10', command=m10)
p100_btn.grid(row=hvps2_level, column=0)
p10_btn.grid(row=hvps2_level, column=1)
m10_btn.grid(row=hvps2_level, column=2)
m100_btn.grid(row=hvps2_level, column=3)

# High voltage readout configuration
vread = 0
def vreadf():
    """Function to update the voltage readout continuously."""
    global vread
    vread = round(adc.readadc() * 1000, 2)
    vread_label1['text'] = vread
    vread_label1.after(300, vreadf)  # Schedule the next readout update

vread_label = ttk.Label(frame1, padding=(20), text='vread (0 ~ 2000 V):', font=('default', '24'))
vread_label1 = ttk.Label(frame1, padding=(30), text='0', font=('default', '54'))
vread_label2 = ttk.Label(frame1, padding=(20), text='V', font=('default', '24'))
vread_label.grid(row=vread_level, column=0)
vread_label1.grid(row=vread_level, column=1, columnspan=2)
vread_label2.grid(row=vread_level, column=3)

# Checkbuttons for operational control
v1 = BooleanVar()
cb1 = ttk.Checkbutton(frame1, padding=(40), text='Dump Open', variable=v1, command=lambda: GPIO.output(gpio_relay[0], v1.get()))
v2 = BooleanVar()
cb2 = ttk.Checkbutton(frame1, padding=(40), text='HVPS Close', variable=v2, command=lambda: GPIO.output(gpio_relay[1], v2.get()))
v3 = BooleanVar()
cb3 = ttk.Checkbutton(frame1, padding=(40), text='HVPS Start', variable=v3, command=lambda: GPIO.output(gpio_relay[2], v3.get()))
v4 = BooleanVar()
cb4 = ttk.Checkbutton(frame1, padding=(40), text='TRG Ready', variable=v4, command=lambda: GPIO.output(gpio_relay[3], v4.get()))
exit_button = ttk.Button(frame1, padding=(50), text='Exit', command=lambda: [dac10v.sethvps(0), time.sleep(1), root.quit(), root.destroy()])
reset_button = ttk.Button(frame1, padding=(50), text='Emergency STOP (RESET)', command=lambda: [reset()])
cb1.grid(row=control_level, column=0)
cb2.grid(row=control_level, column=1)
cb3.grid(row=control_level, column=2)
cb4.grid(row=control_level, column=3)
reset_button.grid(row=exit_level, column=1, columnspan=1)
exit_button.grid(row=exit_level, column=0, columnspan=1)
logo_label.grid(row=exit_level, column=3, columnspan=2)

# Automatic charging control
auto_charge = ttk.Button(frame1, padding=(50), text='Auto Start Charge', command=lambda: [message_charging_set(), time.sleep(waitasec), GPIO.output(gpio_relay[0], 1), v1.set(1), time.sleep(waitasec), GPIO.output(gpio_relay[1], 1), v2.set(1), time.sleep(waitasec), GPIO.output(gpio_relay[2], 1), v3.set(1), time.sleep(waitasec), stopcharge(), trgwait()])
auto_charge_label = ttk.Label(frame1, padding=(50), text='Before charging', font=('default', '32', 'bold'))
auto_charge.grid(row=auto_control_level, column=0, columnspan=1)
auto_charge_label.grid(row=auto_control_level, column=1, columnspan=4)

# Logic function definitions
def stopcharge():
    """Function to stop charging when voltage threshold is reached."""
    global id, charged
    charged = 0
    id = root.after(300, stopcharge)
    if vread > float(hvps) * 0.98 and vread > 10:
        message_beforshot_set()
        time.sleep(0.5)
        GPIO.output(gpio_relay[1], 0)
        v2.set(0)
        time.sleep(waitasec)
        GPIO.output(gpio_relay[2], 0)
        v3.set(0)
        time.sleep(0.5)
        GPIO.output(gpio_relay[3], 1)
        v4.set(1)
        charged = 1
        root.after_cancel(id)

def trgwait():
    """Function to wait for a trigger signal after charging."""
    global charged, idd
    idd = root.after(300, trgwait)
    if charged == 1 and vread < float(hvps) * 0.1:
        reset()
        message_aftershot_set()
        charged = 0
        root.after_cancel(idd)

def reset():
    """Reset all operations and GPIO states."""
    root.after_cancel(idd)
    GPIO.output(gpio_relay[0], 0)
    v1.set(0)
    time.sleep(waitasec)
    GPIO.output(gpio_relay[1], 0)
    v2.set(0)
    time.sleep(waitasec)
    GPIO.output(gpio_relay[2], 0)
    v3.set(0)
    time.sleep(waitasec)
    GPIO.output(gpio_relay[3], 0)
    v4.set(0)
    time.sleep(waitasec)
    message_default_set()

def message_charging_set(): auto_charge_label['text'] = 'Charging !!'
def message_beforshot_set(): auto_charge_label['text'] = 'Waiting for a trigger !!'
def message_default_set(): auto_charge_label['text'] = 'Before charging.'
def message_aftershot_set(): auto_charge_label['text'] = 'You made a shot !'

# Main loop to start the GUI
vreadf()
root.mainloop()

# Cleanup GPIO settings after closing the GUI
for i in range(4):
    GPIO.cleanup(gpio_relay[i])
\end{lstlisting}